\documentclass[usegraphicx,usenatbib,fleqn,useAMS]{mn2e}

\usepackage{amsmath}
\usepackage[varg]{txfonts}
\usepackage{mathrsfs}
\usepackage{braket}
\usepackage{cancel}

\usepackage{CJK}
\usepackage[utf8]{inputenc}
\usepackage[OT2,OT1,T1]{fontenc}
\usepackage[russian,dutch,french,german,english]{babel}

\newcommand\kname{\CJKfamily{mj}안진혁(安振爀)}
\newcommand\cadd{\CJKfamily{gbsn}北京市~朝阳区~大屯路~甲20号，中国科学院~国家天文台}

\newcommand\tens[1]{\ensuremath{\mbox{\boldmath ${\sl #1}$}}}

\renewcommand\pi\upi
\renewcommand\partial\upartial
\renewcommand\le\oldleq
\renewcommand\ge\oldgeq
\newcommand\rmd{\mathrm d}
\newcommand\R{\mathbb R}
\newcommand\be{\bmath e}
\newcommand\he{\hat{\be}}
\newcommand\hbk{\skew{-3}\hat\Bbbk}
\newcommand\del{\bmath\nabla}
\newcommand\dotp{\bmath\cdot}
\newcommand\cross{\bmath\times}
\newcommand\hrpm{\skew{-5}\hat{\bmath{r_\pm}}}
\newcommand\hrmp{\skew{-5}\hat{\bmath{r_\mp}}}
\newcommand\hrp{\skew{-5}\hat{\bmath{r_+}}}
\newcommand\hrm{\skew{-5}\hat{\bmath{r_-}}}
\newcommand\hrg{\skew2\hat{\bmath r}_\gamma}

\title[axisymmetric separable potentials]
{When is an axisymmetric potential separable?}
\author[An]{J.~An\thanks{\kname;~E-mail:~jinan@nao.cas.cn}
\\National Astronomical Observatories, Chinese Academy of Sciences,
A20 Datun Road, Chaoyang District,\thanks{\cadd}
Beijing~100012, PR~China.}

\journal{Mon.~Not.~R.~Astron.~Soc.~{\bf 000},
\pageref{firstpage}--\pageref{lastpage} (2013) {\sc in press}}
\pagerange{\pageref{firstpage}--\pageref{lastpage}}\volume{000}\pubyear{2013}
\date{Accepted 2013 August 7. Received 2013 August 5;
in original form 2013 June 15}

\begin{document}
\label{firstpage}
\begin{CJK*}{UTF8}{}
\maketitle

\begin{abstract}\noindent
An axially symmetric potential $\psi(\rho,z)=\psi(r,\theta)$ is
completely separable if the ratio $\mathfrak s:\mathfrak k$ is constant.
Here $\mathfrak s=r^{-1}\partial_r\partial_\theta(r^2\psi)$ and
$\mathfrak k=\partial_\rho\partial_z\psi$.
If $\beta=\mathfrak s/\mathfrak k$, then the potential admits an integral
of the form $2I=L^2+\beta v_z^2+2\xi$, where $\xi$ is some function
of positions determined by the potential $\psi$. More generally, an
axially symmetric potential respects the third axisymmetric integral of
motion -- in addition to the classical integrals of the Hamiltonian
and the axial component of the angular momentum -- if there exist three
real constants $a,b,c$ (not all simultaneously zero, $a^2+b^2+c^2\ne0$)
such that $a\mathfrak s+b\mathfrak h+c\mathfrak k=0$, where
$\mathfrak h=r^{-1}\partial_\sigma\partial_\tau(r\psi)$ and $(\sigma,\tau)$
is the parabolic coordinate in the meridional plane such that
$\sigma^2=r+z$ and $\tau^2=r-z$.
\end{abstract}

\end{CJK*}

\section{introduction}

Numerical orbit integrations in realistic potentials indicate
that regular orbits are the norms rather than the exceptions
in astrophysics. However, only known exact integrals of motion
besides those due to the Noether theorem are the separation constants
of the Hamilton--Jacobi equation, which are limited to the class of
potentials known as the St\"ackel potential. None the less, behaviours
of regular orbits of most astrophysical interests appear to be
well approximated by those found in the St\"ackel potential \citep{dZ85-or}.
Hence, understanding the St\"ackel potential is not only the first step
to sort out regular orbits but also of practical importance.

Unfortunately, the St\"ackel potential is usually defined to be
the potential separable in an ellipsoidal
(or one of its degenerate limits) coordinate, which hinders
easy understanding and somewhat obscures physical meaning.
\citet{Ly03} on the other hand presented an elementary derivation
of the third integral of motion in axially symmetric potentials
using only vector calculus. His
idea is based on the realization that the kinetic part of
a non-trivial quadratic integral of motion is the scalar
product of the angular momenta about two foci.
%
The angular momentum with respect to $z=a$ on the symmetry axis
is given by $\bmath L_a=(\bmath r-a\hbk)\cross\bmath v=
\bmath L-a(\hbk\cross\bmath v)$,
where $\hbk$ is the unit vector in the direction of the symmetry axis.
Hence, the kinetic part considered by him is equivalent to
$\bmath L_a\dotp\bmath L_{-a}=L^2-a^2(v^2-v_z^2)$, where
$L=\lVert\bmath L\rVert$, $v=\lVert\bmath v\rVert$
and $v_z=\hbk\dotp\bmath v$.
However, $v^2$ is the kinetic part of the Hamiltonian,
$\mathcal H=\frac12v^2+\psi$ and
the assumption is therefore equivalent to existence of an integral
the kinetic part of which is given by $L^2+a^2v_z^2$.

Here we explore the condition under which
there exists such an integral of motion with the kinetic part
of $L^2+\beta v_z^2$.
This approach is conceptually advantageous as imaginary
magnitudes are replaced by negative constants, and thus
unifies the separability in both prolate and oblate spheroidal coordinates
into variation of a single real parameter. More importantly,
this affords us to express the necessary condition for axially symmetric
potentials to admit the third integral \emph{without} prior knowledge on
the constant itself.

In the next section (Sect.~\ref{sec:proof}), we prove that, given
the axially symmetric potential $\psi(\rho,z)=\psi(r,\theta)$
where $(\rho,\phi,z)$ and $(r,\theta,\phi)$ are cylindrical
and spherical polar coordinates, there exists an integral of motion
with its kinetic part given by $L^2+\beta v_z^2$ if (and only if)
the ratio of two mixed partial derivatives,
$r^{-1}\partial_r\partial_\theta(r^2\psi):\partial_\rho\partial_z\psi$,
is constant (which is the same as $\beta$).
An example, namely the \citet{Ku53} disc potential,
is provided in Sect.~\ref{sec:ex}. In Sect.~\ref{sec:sep}, it is shown that
this encompasses the St\"ackel potentials separable in either
prolate or oblate spheroidal coordinate as well as spherical or
cylindrical polar coordinate (the location of the coordinate
origin is assumed to be known).
Discussion on generalization allowing unspecified origin
(on the symmetry axis) and separability in the rotational parabolic
coordinate follows in Sect.~\ref{sec:prb}. In particular, we find that,
by considering the kinetic part of the form
$aL^2-2b(\bmath v\cross\bmath L)\dotp\hbk+cv_z^2$,
an axially symmetric potential $\psi$ admits the third integral if and only
if the null space of three mixed partial derivatives including the earlier two
and the third, $r^{-1}\partial_\sigma\partial_\tau(r\psi)$, is non-trivial
(i.e.\ there exists a constant-coefficient non-trivial
linear combination of the three that is identically zero).
Here $(\sigma,\tau)$ is the parabolic coordinate in the meridional plane.
We conclude in Sect.~\ref{sec:con}
whilst the Appendices provide an overview of the whole subject
concerning separable (St\"ackel) potentials and quadratic integrals,
which puts the present study on proper wider context.

The treatment here is deliberately elementary,
limited to vector calculus and linear algebra
whenever possible, except in Sect.~~\ref{sec:con} and the Appendices. 
The elliptic and parabolic coordinates are introduced
naturally via a modern geometric approach without presumption
on any prior knowledge. By contrast, the Appendices,
which are independent from the main body,
introduce more sophisticated physical and mathematical ideas
throughout.

\section{proof}
\label{sec:proof}

Consider a conservative dynamic system
governed by Newton's laws of motion.
The motion of a unit-mass tracer is determined by
\begin{equation}\label{eq:eom}
\dot{\bmath r}=\bmath v\,;\qquad
\dot{\bmath v}=-\del\psi,
\end{equation}
where $\psi$ is the potential, which is a scalar function of positions.

The evolution of the angular momentum
$\bmath L=\bmath r\cross\bmath v$ is then given by
$\dot{\bmath L}=\cancel{\dot{\bmath r}\cross\bmath v}
+\bmath r\cross\dot{\bmath v}=-\bmath r\cross\del\psi$.
Thus, for 
$L^2=\lVert\bmath L\rVert^2=\bmath L\dotp\bmath L$,
\begin{equation}\label{eq:L2}
\tfrac12D_tL^2
=\bmath L\dotp\dot{\bmath L}
=(\bmath v\cross\bmath r)\dotp\bigl(\bmath r\cross\del\psi\bigr)
=\bmath v\dotp\bigl[\bmath r\cross(\bmath r\cross\del\psi)\bigr],
\end{equation}
where $D_t=\rmd/\rmd t$.
Let $\be^q=\del q$ and $\partial_q=\partial/\partial q$ be
the reciprocal frame vector and the partial differential operator. Then
$\del=\be^r\partial_r+\be^\theta\partial_\theta+\be^\phi\partial_\phi$
in the spherical polar coordinate $(r,\theta,\phi)$ whilst
$\set{\he_r=\be^r,\he_\theta=r\be^\theta,\he_\phi=r\sin\theta\,\be^\phi}$
constitutes the corresponding set of orthonormal frame vectors.
Then $D_tL^2=-2\bmath v\dotp\bmath s$, where
\begin{equation}\label{eq:L2a}\begin{split}
\bmath s&=-\bmath r\cross(\bmath r\cross\del\psi)=
r^2\del\psi-\bmath r(\bmath r\dotp\del\psi)
\\&=r^2\bigl(\del\psi-\be^r\partial_r\psi\bigr)
=\bigl(\be^\theta\partial_\theta
+\be^\phi\partial_\phi\bigr)(r^2\psi).
\end{split}\end{equation}
Here we have used $\bmath r=r\he_r=r\be^r$
and $\be^r\dotp\be^r=\he_r\dotp\he_r=1$.

Meanwhile, $\dot v_z=-\partial_z\psi$ and thus
\begin{equation}\label{eq:vz}
\tfrac12D_tv_z^2
=v_z\dot v_z=-\bmath v\dotp\hbk\,\partial_z\psi,
\end{equation}
where $\hbk$ is the fixed unit vector in the $z$-direction
so that $v_z=\bmath v\dotp\hbk$. The set
$\set{\he_\rho=\he_r\sin\theta+\he_\theta\cos\theta,\he_\phi,\hbk}$
consists in the orthonormal frame vectors for
the cylindrical polar coordinate $(\rho=r\sin\theta,\phi,z=r\cos\theta)$
whilst its reciprocal frame vector set is composed of
$\set{\be^\rho=\he_\rho,\be^\phi=\he_\phi/\rho,\be^z=\hbk}$ and so
$\del=
\he_\rho\partial_\rho+\hbk\,\partial_z+\rho^{-1}\he_\phi\partial_\phi$.

Suppose $2Q=L^2+\beta v_z^2$ with a constant $\beta$.
Then $\dot Q=-\bmath v\dotp\bmath u$, where
\begin{equation}\label{eq:aux}
\bmath u=\be^\theta\partial_\theta(r^2\psi)
+\be^\phi\partial_\phi(r^2\psi)
+\beta\,\hbk\,\partial_z\psi.
\end{equation}
If $\bmath u=\del\xi-\del\cross\bmath A$ is
the Helmholtz decomposition of the vector field $\bmath u$
in equation (\ref{eq:aux}), then
$\dot Q+\bmath v\dotp\del\xi
=\bmath v\dotp(\del\cross\bmath A)
=(\bmath v\cross\del)\dotp\bmath A$.
Consequently, if $\bmath u$ is curl free, then
$I=Q+\xi$ is conserved along the motion of the tracer.
Since $\del\cross\be^q=\del\cross\del q=0$, we have
\begin{equation}\label{eq:curl}\begin{split}
\del\cross\bmath u
&=\be^r\cross\be^\theta\partial_r\partial_\theta(r^2\psi)
+\be^r\cross\be^\phi\partial_r\partial_\phi(r^2\psi)
-\beta\,\hbk\cross\del(\partial_z\psi)
\\&=\he_\phi\left[
\frac1r\frac{\partial^2(r^2\psi)}{\partial r\partial \theta}
-\beta\frac{\partial^2\psi}{\partial\rho\partial z}\right]
-\frac{\he_\theta}\rho\frac{\partial(r^2\psi_\phi)}{\partial r}
+\beta\frac{\he_\rho}\rho\frac{\partial\psi_\phi}{\partial z},
\end{split}\end{equation}
where the partial derivatives with respect to $(r,\theta)$ and $(\rho,z)$
are, respectively, in the spherical and cylindrical polar coordinates
and $\psi_\phi=\partial_\phi\psi$, which is consistently defined
in both coordinates provided that the $z$-axis coincides. Hence, if
$\psi_\phi=0$ and
\begin{equation}\label{eq:rat}
\beta=
\frac{\partial_r\partial_\theta(r^2\psi)}{r\,\partial_\rho\partial_z\psi}
\end{equation}
is constant, then 
there exists a scalar function $\xi$ such that
\begin{equation}\label{eq:qint}
2I=L^2+\beta v_z^2+2\xi
\end{equation}
is an integral of motion and
\begin{equation}\label{eq:dlam}
\del\xi
=r\he_\theta\frac{\partial\psi}{\partial\theta}\biggr\rvert_{r,\phi}
+\beta\,\hbk\frac{\partial\psi}{\partial z}\biggr\rvert_{\rho,\phi}.
\end{equation}
Here the partial derivatives are done with holding the coordinate
variables in the subscripts fixed.

\section{Example}
\label{sec:ex}

Consider the Kuzmin disc \citep{Ku53,Ku56} potential:
\begin{equation}\label{eq:Kd}
\psi=-\frac kb,\quad
b=\sqrt{\rho^2+(a+\lvert z\rvert)^2}
=\sqrt{r^2+2ar\lvert\cos\theta\rvert+a^2},
\end{equation}
which is generated by an infinitesimally thin disc
with the surface density $\Sigma(\rho)\propto(a^2+\rho^2)^{-3/2}$
\citep{To63}. It is easy to find
\begin{equation}
\frac{\partial^2\psi}{\partial\rho\partial z}
=-\frac{3k}{b^5}\rho(a+\lvert z\rvert);\
\frac{\partial^2(r^2\psi)}{\partial r\partial\theta}
=-\frac{3a^2k}{b^5}r^2(a+r\lvert\cos\theta\rvert)\sin\theta,
\end{equation}
and thus the ratio of equation (\ref{eq:rat}) is constant, $\beta=a^2$.
Therefore, the potential of equation (\ref{eq:Kd}) admits an integral
that is quadratic to the velocities \citep[e.g.,][]{dZ85-pe} in the form of
equation (\ref{eq:qint}).

The function $\xi$ is a solution of the partial differential equations
resulting from equation (\ref{eq:dlam}). Using
$r\he_\theta=
z\he_\rho-\rho\,\hbk$
and $\partial_\theta=
z\partial_\rho-\rho\partial_z$,
equation (\ref{eq:dlam}) for equation (\ref{eq:Kd}) with $\beta=a^2$ reduces to
\begin{equation}\begin{split}
\frac{\partial\xi}{\partial\rho}
&=z^2\frac{\partial\psi}{\partial\rho}-\rho z\frac{\partial\psi}{\partial z}
=-\frac{ak\rho\lvert z\rvert}{b^3};\\
\frac{\partial\xi}{\partial z}
&=(a^2+\rho^2)\frac{\partial\psi}{\partial z}-\rho z\frac{\partial\psi}{\partial\rho}
=\frac z{\lvert z\rvert}
\frac{ak}{b^3}\bigl[\rho^2+a(a+\lvert z\rvert)\bigr]
\end{split}\end{equation}
in the cylindrical polar coordinate. Note that the compatibility condition
$\partial_z(\partial_\rho\xi)=\partial_\rho(\partial_z\xi)$
is satisfied (i.e.\ eq.~\ref{eq:curl} being zero), and so
the solution $\xi(\rho,z)$ exists. Let us next suppose that the integral curve
of constant $\xi$ is parametrized by $s$, which then implies
\begin{subequations}
\begin{equation}
\frac\rmd{\rmd s}\xi[\rho(s),z(s)]
=\frac{\partial\xi}{\partial\rho}\frac{\rmd\rho}{\rmd s}
+\frac{\partial\xi}{\partial z}\frac{\rmd z}{\rmd s}=0.
\end{equation}
Hence, the tangential slope of the constant-$\xi$ curve is given by
\begin{equation}
\frac{\rmd\rho}{\rmd z}=\frac{\rho'(s)}{z'(s)}=
-\frac{\partial_z\xi}{\partial_\rho\xi}=
\frac{\rho^2+a(a+\lvert z\rvert)}{\rho z},
\end{equation}
\end{subequations}
which is a first-order ordinary differential equation on $\rho(z)$.
Here we have used $\lvert z\rvert^2=z^2$ for any real $z$.
This is integrated through
\begin{subequations}
\begin{equation}\begin{split}
&\frac12\frac\rmd{\rmd z}\biggl(\frac{\rho^2}{z^2}\biggr)
=\frac\rho{z^2}\frac{\rmd\rho}{\rmd z}-\frac{\rho^2}{z^3}
=\frac{a(a+\lvert z\rvert)}{z^3};
\\&
\frac{\rho^2}{2z^2}=\tilde c-\frac{a^2}{2z^2}-\frac a{\lvert z\rvert}
=\frac c2-\frac{(a+\lvert z\rvert)^2}{2z^2},
\end{split}\end{equation}
where $\tilde c=(c-1)/2$ is an integration constant. Here the last is
the implicit equation of the integral curve. That is to say,
\begin{equation}
F(\rho,z)=\frac{\rho^2+(a+\lvert z\rvert)^2}{z^2}=\frac{b^2}{z^2}=c
\end{equation}
\end{subequations}
coincides with the curve of constant $\xi$ and thus
there exists a real function $f\colon\R\to\R$
such that $\xi(\rho,z)=f[F(\rho,z)]$, which is found using
$\partial_\rho\xi(\rho,z)=f'[F(\rho,z)]\,\partial_\rho F(\rho,z)$. In other words,
\begin{equation}
f'(F)
=\frac{\partial_\rho\xi}{\partial_\rho F}
=-\frac{ak\lvert z\rvert^3}{2b^3}
=-\frac{ak}2\biggl(\frac{z^2}{b^2}\biggr)^{3/2}
=-\frac{ak}{2F^{3/2}}.
\end{equation}
Finally, the antiderivative results in
\begin{equation}\label{eq:lamk}
\xi(\rho,z)=f(F)=\frac{ak}{F^{1/2}}
=\frac{ak\lvert z\rvert}b.
\end{equation}
The integration constant here is immaterial for our purpose
and so set to zero. It is a straightforward task to verify that
equation (\ref{eq:qint}) with $\xi$ given by equation (\ref{eq:lamk})
is indeed a constant of motion for a particle moving under the potential
of equation (\ref{eq:Kd}).

Similarly, if we consider the potential--density profile
given by
\begin{equation}
\psi=-\frac k{\sqrt{(\rho+a)^2+z^2}}
\,;\quad
\frac{\nabla^2\psi}{4\pi G}
=\frac k{4\pi G}\frac a{\rho\bigl[(\rho+a)^2+z^2\bigr]^{3/2}},
\end{equation}
then we find that the ratio in equation (\ref{eq:rat}) is also constant
$\beta=-a^2<0$ (but negative unlike the potential of eq.~\ref{eq:Kd}).
Therefore, this potential too admits a quadratic third integral of
the form of equation (\ref{eq:qint}) with $\beta=-a^2$. The corresponding
function $\xi$ is found similarly, namely
\begin{equation}
\xi=\frac{ka(a+\rho)}{\sqrt{(\rho+a)^2+z^2}}.
\end{equation}

\section{Separable potential}\label{sec:sep}
\subsection{Prolate spheroidal coordinates}

\begin{figure}\begin{center}
\includegraphics[width=\hsize]{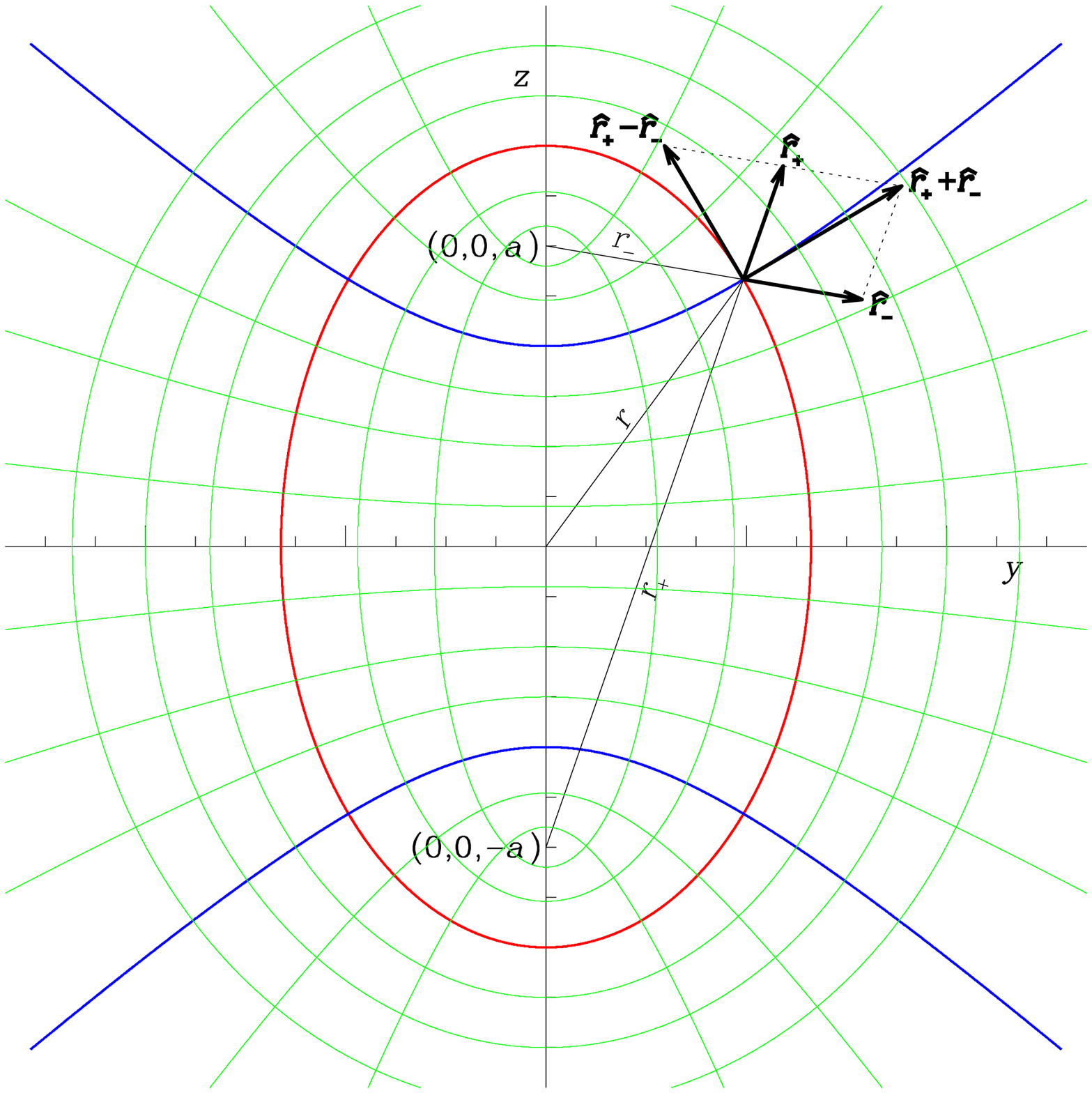}\end{center}
\caption{\label{fig:psc}
$yz$-plane  cross-section of prolate spheroidal coordinate surfaces.}
\end{figure}

Let us consider the quadratic form on $\bmath v$ \citep[cf.][]{Ly03},
\begin{subequations}\begin{equation}
\mathsf{\tilde Q}_a(\bmath v)
=(\bmath{r_+}\cross\bmath v)\dotp(\bmath{r_-}\cross\bmath v),
\end{equation}
where $\bmath{r_\pm}=\bmath r\pm a\hbk$.
Then $\bmath{r_\pm}\cross\bmath v=\bmath L\pm a\bmath\ell$,
where $\bmath\ell=\hbk\cross\bmath v$, and so
$\mathsf{\tilde Q}_a(\bmath v)=L^2-a^2\ell^2$. Yet
$\ell^2=\lVert\bmath\ell\rVert^2=
(\hbk\dotp\hbk)(\bmath v\dotp\bmath v)-(\hbk\dotp\bmath v)^2
=v^2-v_z^2$, and thus $\mathsf{\tilde Q}_a(\bmath v)$
is the kinetic part of an integral of motion,
$\tilde I=2(I-a^2\mathcal H)$ where $I$ is given by equation (\ref{eq:rat})
with a positive constant $\beta=a^2>0$.

The quadratic form $\mathsf{\tilde Q}_a(\bmath v)$ is useful in finding
its diagonalizing frame (in which kinetic parts of $I$ and $\tilde I$
also diagonalize). First note
\begin{equation}
\mathsf{\tilde Q}_a(\bmath v)
=\bigl[(\bmath{r_\pm}\cross\bmath v)\cross\bmath{r_\mp}\bigr]\dotp\bmath v,
\end{equation}\end{subequations}
which further implies
$\mathsf{\tilde Q}_a(\bmath v)=\bmath v\dotp \tens K_a(\bmath v)$ with
\begin{equation}\label{eq:tenk}
2\tens K_a(\bmath v)=(\bmath{r_+}\cross\bmath v)\cross\bmath{r_-}
+(\bmath{r_-}\cross\bmath v)\cross\bmath{r_+},
\end{equation}
which is a self-adjoint linear function of $\bmath v$
(i.e.\ a symmetric tensor). If $\set{\he_1,\he_2,\he_3}$ is
the set of its orthonormal eigenvectors and
$\kappa_i$ is the eigenvalue associated with $\he_i$
such that $\tens K_a(\he_i)=\kappa_i\he_i$,
then $\tens K_a(\bmath v)=\sum_{i=1}^3\kappa_iv_i\he_i$
and $\mathsf{\tilde Q}_a(\bmath v)=\sum_{i=1}^3\kappa_iv_i^2$, where
$v_i=\he_i\dotp\bmath v$ is the velocity component projected on to $\he_i$.
In other words, $\mathsf{\tilde Q}_a$
diagonalizes in the frame consisting of eigenvectors of $\tens K_a$.

These eigenvectors are found by observing
\begin{subequations}
\begin{equation}\begin{split}
2\tens K_a(\bmath{r_\pm})&
=(\bmath{r_\mp}\cross\bmath{r_\pm})\cross\bmath{r_\pm}
=(\bmath{r_+}\dotp\bmath{r_-})\bmath{r_\pm}-r_\pm^2\bmath{r_\mp};
\\2\tens K_a(\hrpm)
&=(\bmath{r_+}\dotp\bmath{r_-})\hrpm-r_+r_-\hrmp,
\end{split}\end{equation}
where
$r_\pm=\lVert\bmath{r_\pm}\rVert=\sqrt{\bmath{r_\pm}\dotp\bmath{r_\pm}}$
and $\hrpm=\bmath{r_\pm}/r_\pm$. Hence,
\begin{equation}
2\tens K_a(\hrp\pm\hrm)
=\bigl(\bmath{r_+}\dotp\bmath{r_-}\mp r_+r_-)(\hrp\pm\hrm),
\end{equation}
that is, $\hrp\pm\hrm$ is the two of
(orthogonal but not necessarily normalized)
eigenvectors of $\tens K_a$.
The remaining eigenvector is $\he_\phi$, which is,
up to signs, the only unit vector orthogonal to the first two.
The associated eigenvalue is found by direct calculations, namely,
%
$\tens K_a(\he_\phi)=(r^2-a^2)\he_\phi$.
\end{subequations}
It follows that
if $v_\pm$ is the velocity component in the direction of $\hrp\pm\hrm$,
then the diagonalized $\mathsf{\tilde Q}_a(\bmath v)$ is given by
\begin{equation}
2\mathsf{\tilde Q}_a(\bmath v)
=(l-m)\,v_+^2+(l+m)\,v_-^2+2lv_\phi^2,
\end{equation}
where $m= r_+r_-$ and $l=\bmath{r_+}\dotp\bmath{r_-}=r^2-a^2$.

In order to interpret these eigenvectors geometrically, we note
\begin{equation}\begin{split}
&\bmath{r_\pm}=\bmath r\pm a\,\hbk
=\rho\he_\rho+(z\pm a)\,\hbk,\quad
r_\pm^2=\rho^2+(z\pm a)^2;
\\&
\del r_\pm^2=2\rho\be^\rho+2(z\pm a)\be^z=2\bmath{r_\pm},\quad
\del r_\pm^2=2r_\pm(\del r_\pm);
\\&\therefore
\hrpm=\bmath{r_\pm}/r_\pm=\del r_\pm.
\label{eq:hatr}\end{split}\end{equation}
Hence $\hrp\pm\hrm$ is normal to the surface defined by
a constant value of $r_+\pm r_-$, respectively.
Since $r_\pm=\lVert\bmath r\pm a\,\hbk\rVert$
is the respective distance to the point $z=\mp a$ on the $z$-axis,
the constant sum defines a confocal set of prolate spheroids
whilst the constant difference does that of circular hyperboloids
of two sheets -- see Fig.~\ref{fig:psc}.
Together with the meridional planes, they constitute
a pair-wise orthogonal foliation of the three-dimensional
Euclidean space ($\mathbb R^3$). Any set of monotonic differentiable functions
on each leaf in the family then defines an orthogonal coordinate
in which $\mathsf{\tilde Q}_a$ diagonalizes.
Within arbitrary scalings, this coordinate system corresponds to the prolate
spheroidal coordinate with the foci on $(x,y,z)=(0,0,\pm a)$.
An intuitive choice for the meridional coordinate variables is
$(\zeta,\eta)=(r_++r_-,r_+-r_-)$, that is,
the sum of and the difference between the distances to two foci.
Here $\zeta/2$ and $\eta/2$ are also the semimajor axes of the meridional
ellipse and hyperbola, and so each essentially labels
the particular coordinate surface. More common choice however is either
$(\upsilon,\nu,\phi)$, where $(\zeta,\eta)=(2a\cosh\upsilon,2a\cos\nu)$
\citep[e.g.,][]{KdZ91,Bi12} or $(\lambda,\mu,\phi)$, where
$(\zeta^2,\eta^2)=[4(\lambda+a^2+\alpha),4(\mu+a^2+\alpha)]$
with a constant $\alpha$ \cite[e.g.,][]{ddZ88,Sa12}.

The potential admitting the integral $\tilde I$
is characterized in the same 
coordinate system by considering the vector field,
\begin{subequations}
\begin{equation}
2\tens K_a(\del\psi)=-\bigl[\bmath{r_+}\cross(\bmath{r_-}\cross\del\psi)+
\bmath{r_-}\cross(\bmath{r_+}\cross\del\psi)\bigr].
\end{equation}
Since $\bmath{r_\pm}\cross\del\psi=
\he_\phi\bigl(\partial_\theta\psi\rvert_r
\pm a\partial_\rho\psi\rvert_z\bigr)$, we find (cf.\ eq.~\ref{eq:dlam})
\begin{equation}
\tens K_a(\del\psi)=r\he_\theta\partial_\theta\psi\rvert_r
-a^2\he_\rho\partial_\rho\psi\rvert_z=\del(\xi-a^2\psi),
\end{equation}
\end{subequations}
and so $\tens K_a(\del\psi)$ is curl free.
For the meridional coordinate
$(\zeta,\eta)=(r_++r_-,r_+-r_-)$,
equation (\ref{eq:hatr}) indicates that
$(\del\zeta,\del\eta)=(\hrp+\hrm,\hrp-\hrm)$ and
$\del=(\hrp+\hrm)\partial_\zeta+(\hrp-\hrm)\partial_\eta
+\be^\phi\partial_\phi$.
Defining $\mathcal D_\pm=\partial_\zeta\pm\partial_\eta$
and also assuming $\partial_\phi=0$, we then have
\begin{gather}
\hrp\cross\del=\hrp\cross\hrm\mathcal D_-,\quad
\hrm\cross\del=\hrm\cross\hrp\mathcal D_+;\nonumber\\
\bmath{r_\pm}\cross(\bmath{r_\mp}\cross\del\psi)=
\bmath{r_\pm}\cross(\bmath{r_\mp}\cross\hrpm)\mathcal D_\pm(\psi)
=(m\hrmp-l\hrpm)
\mathcal D_\pm(\psi);
\nonumber\\\label{eq:rcur}
\tens K_a(\del\psi)=\bigl[\hrp\,(l\mathcal D_+-m\mathcal D_-)
+\hrm\,(l\mathcal D_--m\mathcal D_+)\bigr]\,\psi,
\end{gather}
where again $m=r_+r_-$ and $l=\bmath{r_+}\dotp\bmath{r_-}=r^2-a^2$.
Consequently
%
\begin{equation}
\del\cross\tens K_a(\del\psi)
=2a\rho\he_\phi\bigl(\mathcal D_-^2-\mathcal D_+^2\bigr)\,(m\psi),
\end{equation}
provided that $\partial_\zeta\partial_\eta=\partial_\eta\partial_\zeta$.
Calculations ease using $\mathcal D_\pm(r_\pm)=1$ and $\mathcal D_\mp(r_\pm)=0$,
which follows $2r_\pm=\zeta\pm\eta$. In addition, $\mathcal D_\pm(l)=r_\pm$
from $\zeta^2+\eta^2=2(r_+^2+r_-^2)=4(r^2+a^2)$ and so $l=(r_+^2+r_-^2)/2-2a^2$.
Thus $\del\cross\tens K_a(\del\psi)=0$ is equivalent to
\begin{subequations}\label{eq:stk}
\begin{equation}
\bigl(\mathcal D_+^2-\mathcal D_-^2\bigr)\,(m\psi)=
4\frac{\partial^2(m\psi)}{\partial\zeta\partial\eta}=0
%
\quad\Rightarrow\
\psi=\frac{f(\zeta)+g(\eta)}m,
\end{equation}
where $f$ and $g$ are arbitrary functions of respective argument and
\begin{equation}\begin{split}
m&=r_+r_-=(\zeta^2-\eta^2)/4=\lambda-\mu
\\&=a^2(\cosh^2\upsilon-\cos^2\nu)
=a^2(\sinh^2\upsilon+\sin^2\nu)
\end{split}\end{equation}
\end{subequations}
with $(\upsilon,\nu)$ and $(\lambda,\mu)$ as defined earlier.
Hence, an axially symmetric potential with the ratio in equation (\ref{eq:rat})
being a positive constant $\beta=a^2>0$ is the St\"ackel potential
\citep[e.g.,][]{Ly62,dZ85-kt,EL89} separable
in the prolate spheroidal coordinate with the foci
located on $z=\pm a$ on the symmetry axis.

The opposite implication, that any St\"ackel potential separable in
a prolate spheroidal coordinate results in a positive constant for
equation (\ref{eq:rat}), is trivial from its transformations
back to the cylindrical and spherical polar coordinates.
For instance, with the coordinate transform
$(\lambda,\mu)\to(\rho^2,z^2)$ \citep[see e.g.,][]{Ly62},
\begin{subequations}
\begin{equation}\label{eq:sphct}
\rho^2=\frac{(\lambda+\alpha)(\mu+\alpha)}{\alpha-\beta},\quad
z^2=\frac{(\lambda+\beta)(\mu+\beta)}{\beta-\alpha},
\end{equation}
the differential equation in equation (\ref{eq:stk}) transforms to
\begin{multline}
\frac{\partial^2\bigl[(\lambda-\mu)\psi\bigr]}{\partial\lambda\partial\mu}
=\frac{\partial\psi}{\partial\mu}
-\frac{\partial\psi}{\partial\lambda}
+(\lambda-\mu)\frac{\partial^2\psi}{\partial\lambda\partial\mu}
=\frac{\lambda-\mu}{4(\beta-\alpha)}
\\\times\Biggl[\frac1{z^3}\frac\partial{\partial z}
\biggl(z^3\frac{\partial\psi}{\partial z}\biggr)
-\frac1{\rho^3}\frac\partial{\partial\rho}
\biggl(\rho^3\frac{\partial\psi}{\partial\rho}\biggr)
+\frac{\rho^2-z^2+\beta-\alpha}{\rho z}
\frac{\partial^2\psi}{\partial\rho\partial z}\Biggr]
\end{multline}
\end{subequations}
in the cylindrical polar coordinate. We furthermore find that
\begin{subequations}
\begin{multline}
\frac1{\rho^3}\frac\partial{\partial\rho}
\biggl(\rho^3\frac{\partial\psi}{\partial\rho}\biggr)
-\frac1{z^3}\frac\partial{\partial z}
\biggl(z^3\frac{\partial\psi}{\partial z}\biggr)
-\frac{\rho^2-z^2}{\rho z}
\frac{\partial^2\psi}{\partial\rho\partial z}
\\=\biggl(\frac1{\rho^kz^{3-k}}\frac\partial{\partial\rho}
+\frac1{\rho^{1+k}z^{2-k}}\frac\partial{\partial z}\biggr)\,
\biggl(\rho^kz^{3-k}\frac{\partial\psi}{\partial\rho}
-\rho^{1+k}z^{2-k}\frac{\partial\psi}{\partial z}\biggr),
\end{multline}
where $k$ is any fixed real number. However,
the transformation to the spherical polar coordinate
$\partial_\rho=\sin\theta\,\partial_r+r^{-1}\!\cos\theta\,\partial_\theta$
and $\partial_z=\cos\theta\,\partial_r-r^{-1}\!\sin\theta\,\partial_\theta$
indicates that
\begin{equation}\begin{split}
\frac1{\rho^kz^{3-k}}&\frac\partial{\partial\rho}
+\frac1{\rho^{1+k}z^{2-k}}\frac\partial{\partial z}
=\frac1{r^3\sin^{1+k}\!\theta\cos^{3-k}\!\theta}
\frac\partial{\partial r},
\\\rho^kz^{3-k}&\frac\partial{\partial\rho}
-\rho^{1+k}z^{2-k}\frac\partial{\partial z}
=r^2\sin^k\!\theta\cos^{2-k}\!\theta\frac\partial{\partial\theta}.
\end{split}\end{equation}
\end{subequations}
So the St\"ackel potential separable in a prolate
spheroidal coordinate satisfies the differential equation
\citep[cf.][eq.~9]{Sa12}
\begin{equation}
\frac{\partial^2\bigl[(\lambda-\mu)\psi\bigr]}{\partial\lambda\partial\mu}
=\frac{\lambda-\mu}{4\rho z\delta}
\Biggl[\delta
\frac{\partial^2\psi}{\partial\rho\partial z}
-\frac1r\frac\partial{\partial r}\biggl(
r^2\frac{\partial\psi}{\partial\theta}\biggr)\Biggr]=0,
\end{equation}
where $\delta=\beta-\alpha$, which is a positive constant.

\subsection{Oblate spheroidal coordinates}

If $\beta=-a^2<0$ in equation (\ref{eq:rat}) is a negative constant,
the integral of motion $I$ in equation (\ref{eq:qint}) may be written down
to be
%
$2I=\mathsf{\bar Q}_a(\bmath v)
+a^2(\he_\phi\dotp\bmath v)^2+\xi(\bmath r)$
%
introducing another quadratic form:
\begin{equation}
\mathsf{\bar Q}_a(\bmath v)
=(\bmath{q_+}\cross\bmath v)\dotp(\bmath{q_-}\cross\bmath v),
\end{equation}
%
where $\bmath{q_\pm}=\bmath r\pm a\he_\rho$.
Essentially verbatim calculations as in the preceding section
establish that $\mathsf{\bar Q}_a(\bmath v)$ diagonalizes
in the frame defined by
$\set{\skew{-6}\hat{\bmath{q_+}}\pm\skew{-6}\hat{\bmath{q_-}},\he_\phi}$,
where
$\skew{-6}\hat{\bmath{q_\pm}}=\bmath{q_\pm}/\lVert\bmath{q_\pm}\rVert$.
So does the kinetic part of $I$
since $v_\phi^2$ trivially diagonalizes in the same frame.

The meridional cross-section of the coordinate surfaces defined by
this frame set is basically the same as before except for the $R$-
and $z$-axes switched (see again Fig.~\ref{fig:psc}).
That is to say, the symmetry axis now lies along
the minor axis of the ellipse. These result in the coordinate
surfaces being the confocal oblate spheroids and circular hyperboloids
of one sheet whilst the resulting coordinate system is identified
with the oblate spheroidal coordinate with the meridional foci
on the mid-plane point $R=(\pm)a$. Note that
the derivation of equation (\ref{eq:stk}) in the preceding section
considers only the two-dimensional coordinates
restricted to the meridional plane. Therefore, the same calculations
(with the $\bmath{r_\pm}\to\bmath{q_\pm}$ replacement) can be used
to show that the axially symmetric potential that results in
a negative constant $\beta=-a^2$ in equation (\ref{eq:rat}) is also
the St\"ackel potential but separable in an oblate spheroidal coordinate.

\subsection{Degenerate cases}

If $\partial_r\partial_\theta(r^2\psi)=0$ or $\partial_\rho\partial_z\psi=0$,
equation (\ref{eq:rat}) is also considered constant. In fact,
this indicates that the axially symmetric potential is
of the separable form in the spherical or cylindrical polar coordinate, 
\begin{align}
&\partial_r\partial_\theta(r^2\psi)=0
&\Longleftrightarrow&&
\psi&=f(r)+r^{-2}g(\theta);
\label{eq:sph}\\\label{eq:cyl}
&\partial_\rho\partial_z\psi=0
&\Longleftrightarrow&&
\psi&=f(\rho)+g(z).
\end{align}
%
Given arbitrary functions $f$ and $g$,
the third integrals besides the energy, $\mathcal H$, and 
the axial angular momentum component, $L_z$, are,
$\frac12L^2+g(\theta)$ and $\frac12v_z^2+g(z)$, respectively.
These are consistent with equations (\ref{eq:qint}) and (\ref{eq:dlam}),
either setting $\beta=0$ or considering
$\lim_{\beta\rightarrow\infty}\beta^{-1}I$.
With equation (\ref{eq:cyl}), the $z$-motion decouples
and the third integral is basically the corresponding one-dimensional
energy. The third integral for equation (\ref{eq:sph})
on the other hand is essentially the Hamiltonian of the
angular motion projected along the radial direction on to the unit sphere.
If $g=0$ in either case,
the quadratic integral is simply the square of
momentum, namely $\bmath L$ or $v_z$. Note that equation (\ref{eq:sph})
then reduces to spherically
symmetric potentials and is actually superintegrable since $\bmath L$
counts as two additional integrals, $(L_x,L_y)$.

If both $\partial_r\partial_\theta(r^2\psi)=0$
and $\partial_\rho\partial_z\psi=0$, the potential is completely
specified. That is to say, the general solution is then given by
\begin{equation}\label{eq:ham}
\psi=k_0r^2+k_1\rho^{-2}+k_2z^{-2}+k_3,
\end{equation}
which involves constants but not arbitrary functions.
The particular example includes the harmonic potential $(\psi=k_0r^2)$.
This potential is separable in both the spherical and cylindrical
polar coordinates as well as any prolate or oblate spheroidal coordinates
with arbitrary parameter $a$ (formally $\beta=0/0$ is an indeterminate constant).
Dynamics in the potential of equation (\ref{eq:ham}) is not only
superintegrable (i.e.\ the orbit projected on to the meridional
plane becomes a closed curve) but also completely soluble
in a closed form using only elementary functions.
If $k_1=0$, the potential is further separable in
the Cartesian coordinate and so maximally superintegrable
(i.e.\ the orbits are thus truly periodic).

\section{Separability with an unspecified origin}
\label{sec:prb}

The discussion so far has implicitly assumed that the coordinate
origin of the frame about which the angular momentum is defined
is known a priori. Relaxing this restriction is equivalent to allowing
the coordinate origin to be an arbitrary point on the symmetry axis. 
The angular momentum with respect to the point $z=\alpha$ on
the symmetry axis is given by
$\bmath L_\alpha=(\bmath r-\alpha\hbk)\cross\bmath v
=\bmath L-\alpha\bmath\ell$, where $\bmath L$ is
the angular momentum with respect to $z=0$ and 
$\bmath\ell=\hbk\cross\bmath v$. Then
$L_\alpha^2=L^2-2\alpha\omega+\alpha^2\ell^2$, where
$\omega=\bmath\ell\dotp\bmath L=(\bmath v\cross\bmath L)\dotp\hbk$
and $\ell^2=v^2-v_z^2$.
From equation (\ref{eq:eom}),
$\dot{\bmath\ell}=-\hbk\cross\del\psi$ and so 
\begin{subequations}
\begin{gather}
\tfrac12D_t\ell^2
=\bmath\ell\dotp\dot{\bmath\ell}
=(\bmath v\cross\hbk)\dotp(\hbk\cross\del\psi)
=\bmath v\dotp\bigl[\hbk\cross(\hbk\cross\del\psi)\bigr],
\\\begin{split}
\dot\omega=\dot{\bmath\ell}\dotp\bmath L+\bmath\ell\dotp\dot{\bmath L}
&=(\bmath v\cross\bmath r)\dotp(\hbk\cross\del\psi)
+(\bmath v\cross\hbk)\dotp(\bmath r\cross\del\psi)
\\&=\bmath v\dotp\bigl[\bmath r\cross(\hbk\cross\del\psi)
+\hbk\cross(\bmath r\cross\del\psi)\bigr].
\end{split}\end{gather}
\end{subequations}
Hence, $D_tL_\alpha^2=-2\bmath v\dotp(\bmath s-\alpha\bmath h+\alpha^2\bmath k)$,
where $\bmath s$ is as in equation (\ref{eq:L2a}),
\begin{subequations}
\begin{equation}
\bmath s
=-\bmath r\cross(\bmath r\cross\del\psi)
=r^2(\del\psi-\be^r\partial_r\psi),
\end{equation}
and also introduced are the vector fields
\begin{equation}\begin{split}
\bmath h&=-\bigl[\bmath r\cross(\hbk\cross\del\psi)
+\hbk\cross(\bmath r\cross\del\psi)\bigr],
\\\bmath k&=-\hbk\cross(\hbk\cross\del\psi)
=\del\psi-\hbk\,\partial_z\psi.
\end{split}\end{equation}
\end{subequations}

Following a similar argument as in Sect.~\ref{sec:proof}, we conclude that,
if there exist real constants $a,b,c$ such that
(NB: $\del\cross\del\psi=0$)
\begin{equation}\label{eq:cdep}
a\,\del\cross\bmath s-b\,\del\cross\bmath h-c\,\del\cross\bmath k=0
\qquad(a^2+b^2+c^2\ne0),
\end{equation}
then there exists the scalar function $\xi$ which is the solution of
\begin{equation}
\del\xi=a\bmath s-b\bmath h-c\bmath k,
\end{equation}
and therefore the potential admits an integral of motion,
\begin{subequations}
\begin{equation}
2I=aL^2-2b\omega-c\ell^2+2\xi,
\end{equation}
or equivalently
\begin{equation}
2\tilde I=2(I+c\mathcal H)=aL^2-2b\omega+cv_z^2+2(\xi+c\psi).
\end{equation}
\end{subequations}
If $a\ne0$, we can find an integral with the kinetic part
given by $2Q_a=L_\alpha^2+\beta v_z^2$, namely
$a^{-1}I+\beta\mathcal H=a^{-1}\tilde I+\alpha^2\mathcal H
=Q_a+a^{-1}\xi+\beta\psi$,
where $\alpha=b/a$ and $\beta=\alpha^2+c/a$. This indicates that,
depending on $\beta$,
the potential is separable in the prolate spheroidal ($\beta>0$),
the spherical polar ($\beta=0$) or the oblate spheroidal ($\beta<0$)
coordinate with the origin displaced to $z=\alpha$
on the symmetry axis.

For an axisymmetric system ($\partial_\phi=0$),
\begin{equation}\begin{split}
\bmath s&=\be^\theta\partial_\theta(r^2\psi),
&\del\cross\bmath s&=\he_\phi\mathfrak s,
&\mathfrak s&=r^{-1}\partial_r\partial_\theta(r^2\psi);
\\\bmath k&=\be^\rho\partial_\rho\psi,
&\del\cross\bmath k&=\he_\phi\mathfrak k,
&\mathfrak k&=\partial_z\partial_\rho\psi.
\end{split}\end{equation}
However, the vector field $\del\cross\bmath h=\he_\phi\mathfrak h$,
whilst still parallel to $\he_\phi$,
results in rather complicated expressions,
($\psi_r=\partial_r\psi$ \& $\psi_\theta=\partial_\theta\psi$)
\begin{multline}\label{eq:hfrak}
\mathfrak h=\rho(\partial_\rho^2-\partial_z^2)\psi
+3\partial_\rho\psi+2z\partial_z\partial_\rho\psi
\\=r\sin\theta\,\partial_r^2\psi
+2(\sin\theta+\cos\theta\,\partial_\theta)\psi_r
+r^{-1}(\cos\theta-\sin\theta\,\partial_\theta)\psi_\theta
\\=\frac{\sin\theta}r\frac{\partial(r^2\psi_r)}{\partial r}
-\frac1{r\sin^2\!\theta}
\frac{\partial(\psi_\theta\sin^3\!\theta)}{\partial\theta}
+\frac{2\cos\theta}{r^2}
\frac{\partial^2(r^2\psi)}{\partial r\partial\theta}
\end{multline}
in the cylindrical and spherical polar coordinates (NB: the
simplest coordinate expression is obtained with
the rotational parabolic coordinate as shall be shown).
The condition in equation (\ref{eq:cdep})
for an axisymmetric potential is then equivalent to
three functions $\mathfrak{s,h,k}$ being linearly dependent.
Here the linear dependence is considered within
the infinite-dimensional functional space,
not in the sense of the vector field on the three-dimensional
configuration space. The algebraic necessary (but not sufficient)
condition is for all the generalized Wronskians
$\set{W(\mathfrak s,\mathfrak h,\mathfrak k)}$ to identically vanish
and also
\begin{equation}
\begin{vmatrix}
\mathfrak s&\mathfrak h&\mathfrak k\\
\partial_1\mathfrak s&\partial_1\mathfrak h&\partial_1\mathfrak k\\
\partial_2\mathfrak s&\partial_2\mathfrak h&\partial_2\mathfrak k
\end{vmatrix}=0,
\end{equation}
where $\partial_i=\partial/\partial x_i$ and $(x_1,x_2)$ represents
any coordinate on the meridional plane -- e.g., $(\rho,z)$ or $(r,\theta)$.

\subsection{Rotational parabolic coordinates}

\begin{figure}\begin{center}
\includegraphics[width=\hsize]{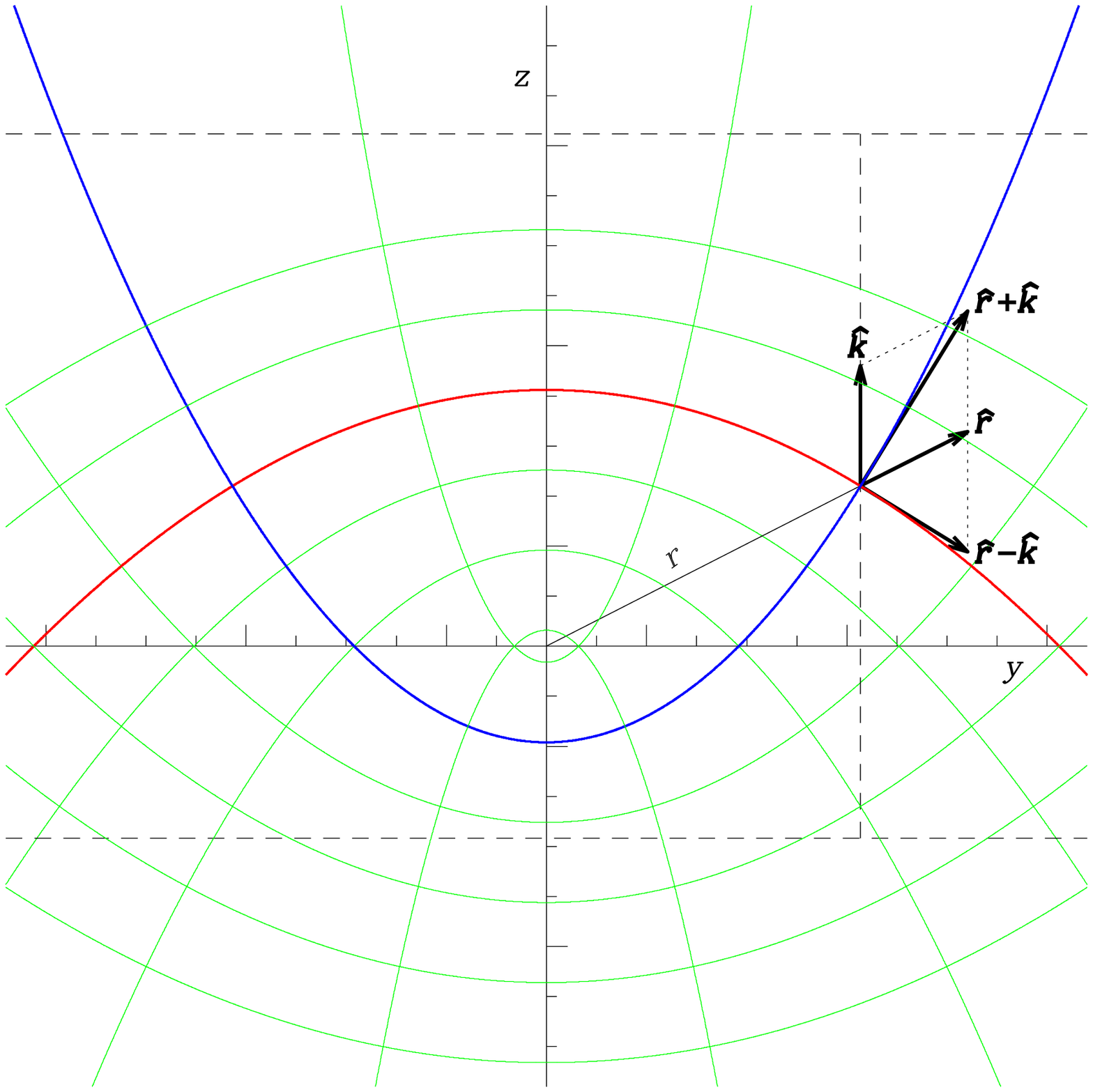}\end{center}
\caption{\label{fig:rpc}
$yz$-plane  cross-section of rotational parabolic coordinate surfaces.}
\end{figure}

If $\mathfrak h/\mathfrak k=2\gamma$ is constant, then
$0\cdot\mathfrak s+1\cdot\mathfrak h-2\gamma\mathfrak k=0$ and so
\begin{equation}\label{eq:pint}
I=\omega-\gamma\ell^2+\xi
\qquad(\del\xi=\bmath h-2\gamma\bmath k)
\end{equation}
is an integral of motion. Here the kinetic part is given by
\begin{equation}
\omega-\gamma\ell^2
=\bmath\ell\dotp(\bmath L-\gamma\bmath\ell)
=\bmath\ell\dotp(\bmath r_\gamma\cross\bmath v)
=\bmath v\dotp\tens P_\gamma(\bmath v),
\end{equation}
where $\bmath r_\gamma=\bmath r-\gamma\hbk$ and
\begin{equation}
2\tens P_\gamma(\bmath v)=-\bigl[\bmath r_\gamma\cross(\hbk\cross\bmath v)
+\hbk\cross(\bmath r_\gamma\cross\bmath v)\bigr]
\end{equation}
is a self-adjoint linear function of $\bmath v$. Next we find
\begin{subequations}
\begin{equation}\label{eq:pcal}\begin{split}
2\tens P_\gamma&(\bmath r_\gamma)=
-\bmath r_\gamma\cross(\hbk\cross\bmath r_\gamma)=
(\bmath r_\gamma\dotp\hbk)\,\bmath r_\gamma-r_\gamma^2\,\hbk,
\\2\tens P_\gamma&(\hbk)=
-\hbk\cross(\bmath r_\gamma\cross\hbk)=
(\bmath r_\gamma\dotp\hbk)\,\hbk-\bmath r_\gamma,
\end{split}\end{equation}
where $r_\gamma=\lVert\bmath r_\gamma\rVert$ is the distance to
$(x,y,z)=(0,0,\gamma)$. Then
\begin{equation}\begin{split}
2\tens P_\gamma(\hrg\pm\hbk)&=
(\bmath r_\gamma\dotp\hbk)\,\hrg-r_\gamma\,\hbk
\pm[(\bmath r_\gamma\dotp\hbk)\,\hbk-r_\gamma\,\hrg]
\\&=(\bmath r_\gamma\dotp\hbk\mp r_\gamma)(\hrg\pm\hbk),
\end{split}\end{equation}
\end{subequations}
where $\hrg=\bmath r_\gamma/r_\gamma$. That is,
$\hrg\pm\hbk$ is the two of eigenvectors of $\tens P_\gamma$.
Here we note $\hrg=\del r_\gamma$ (cf.\ eq.~\ref{eq:hatr})
whilst $\del z=\hbk$. Hence, two eigenvectors are, respectively, normal
to the surfaces defined by constant values of $c_\pm=r_\gamma\pm z>0$.
The geometry of conic sections indicates that the meridional
cross-sections of these surfaces are confocal parabolae with the
focus at $z=\gamma$ on the symmetry axis (which is also the axis
of symmetry of all parabolae) and the directrix given by the
horizontal line of $z=\pm c_\pm$ (see Fig.~\ref{fig:rpc}).
These eigenvectors thus define a pair of orthogonal foliations
consisting of the set of paraboloids of revolution.
They, together with the meridional planes, constitute
the complete set of coordinate surfaces of
the rotational parabolic (or circular paraboloidal)
coordinate with the origin at $(x,y,z)=(0,0,\gamma)$.
The standard choice for the scaling functions of the coordinate variables
is given by $(\sigma,\tau,\phi)$, where
$\sigma^2=r_\gamma+z_\gamma$ and $\tau^2=r_\gamma-z_\gamma$ with
$z_\gamma=z-\gamma$ and $r_\gamma^2=\rho^2+z_\gamma^2$. The inverse
transformation is then $\rho=\sigma\tau$ and
$2z_\gamma=\sigma^2-\tau^2$ (and $2r_\gamma=\sigma^2+\tau^2$).
However, alternative choices of
coordinate variables such that $(\zeta,\eta)=(\sigma^2,\tau^2)$
\citep[e.g.,][who used $\eta=-\tau^2$]{LL,ST97} are not uncommon.

The condition $\mathfrak h/\mathfrak k=2\gamma$ may be expressed in
the parabolic coordinate via
explicit coordinate transformations, but utilizing the tensor $\tens P_\gamma$
simplifies calculations. First we have
$2\tens P_\gamma(\del\psi)=\bmath h-2\gamma\bmath k$,
and so $2\del\cross\tens P_\gamma(\del\psi)
=\he_\phi(\mathfrak h-2\gamma\mathfrak k)$. Hence,
$\mathfrak h=2\gamma\mathfrak k$ is equivalent to
$\tens P_\gamma(\del\psi)$ being curl free.
Equation (\ref{eq:pcal}) together with
$\del=\hrg\partial_{r_\gamma}
+\hbk\partial_{z_\gamma}+\be^\phi\partial_\phi$ then indicates
(NB: $\bmath r_\gamma\dotp\hbk=z_\gamma$)
\begin{subequations}\begin{equation}\begin{split}
2\tens P_\gamma(\del\psi)
&=(z_\gamma\hrg-r_\gamma\hbk)\partial_{r_\gamma}\psi
+(z_\gamma\hbk-\bmath r_\gamma)\partial_{z_\gamma}\psi
\\&=\hrg
(z_\gamma\partial_{r_\gamma}-r_\gamma\partial_{z_\gamma})\psi
-\hbk\,(r_\gamma\partial_{r_\gamma}-z_\gamma\partial_{z_\gamma})\psi.
\end{split}\end{equation}
Next using the identity
$\del\cross(a\bmath v)=(\del a)\cross\bmath v+a(\del\cross\bmath v$),
\begin{equation}\begin{split}
2\del\cross&\tens P_\gamma(\del\psi)
\\&=\hbk\cross\hrg
[\partial_{z_\gamma}(z_\gamma\partial_{r_\gamma}-r_\gamma\partial_{z_\gamma})
+\partial_{r_\gamma}(r_\gamma\partial_{r_\gamma}-z_\gamma\partial_{z_\gamma})]\psi
\\&=r_\gamma^{-1}\rho\he_\phi\,
(\partial_{r_\gamma}^2-\partial_{z_\gamma}^2)(r_\gamma\psi)
=r_\gamma^{-1}\he_\phi\,\partial_\sigma\partial_\tau(r_\gamma\psi),
\end{split}\end{equation}\end{subequations}
where $\partial_\sigma$ and $\partial_\tau$ are the coordinate
partial derivative with respect to
$(\sigma,\tau)=(\sqrt{r_\gamma+z_\gamma},\sqrt{r_\gamma-z_\gamma})$.
Also used are $\del\cross\hrg=\del\cross\hbk=0$
and $\bmath r_\gamma=r_\gamma\hrg
=\rho\he_\rho+z\hbk$, which results in
$r_\gamma\hbk\cross\hrg
=\rho\hbk\cross\he_\rho=\rho\he_\phi$.
Finally, the general solution of $\del\cross\tens P_\gamma(\del\psi)=0$
is then given by
\begin{equation}\label{eq:ppara}
\psi=\frac{f(r_\gamma+z_\gamma)+g(r_\gamma-z_\gamma)}{r_\gamma}
=\frac{\tilde f(\sigma)+\tilde g(\tau)}{\sigma^2+\tau^2},
\end{equation}
where $\tilde f(x)=2f(x^2)$ and $\tilde g(x)=2g(x^2)$. This is simply
any potential separable in the rotational parabolic coordinate
with the origin at $(x,y,z)=(0,0,\gamma)$ \citep[e.g.,][]{ST97}.
In other words, $\mathfrak h/\mathfrak k=2\gamma$ is constant
if and only if $\psi$ is the St\"ackel potential separable in
a rotational parabolic coordinate with the origin located at a point
on the symmetry axis ($z=\gamma$ in particular).
The resulting integral is in the form of
$I=\bmath\ell\dotp\bmath L_\gamma+\xi$, where
$\xi$ is found by integrating $\del\xi=2\tens P_\gamma(\del\psi)$.
In particular
(here $c_\pm=r_\gamma\pm z_\gamma$),
\begin{equation}\label{eq:xpsep}
\xi=\frac{c_+\,g(c_-)-c_-\,f(c_+)}{r_\gamma}
=\frac{\sigma^2\tilde g(\tau)-\tau^2\tilde f(\sigma)}{\sigma^2+\tau^2}.
\end{equation}

If $\mathfrak h=0$, then $\mathfrak h/\mathfrak k=0$.
Consequently,
the potential with $\mathfrak h=0$ is separable in the rotational
parabolic coordinate with the coordinate origin at $(x,y,z)=(0,0,0)$.
In addition, this also implies that given the parabolic
coordinate $(\sigma,\tau,\phi)$ with $\gamma=0$, namely
$\sigma^2=r+z$ and $\tau^2=r-z$, the field $\mathfrak h$
is expressible to be
\begin{equation}
\mathfrak h=\he_\phi\dotp\bigl[2\del\cross\tens P_0(\del\psi)\bigr]
=\frac1{\sigma^2+\tau^2}
\frac{\partial^2\bigl[(\sigma^2+\tau^2)\psi\bigr]}
{\partial\sigma\partial\tau},
\end{equation}
in the same coordinate -- note
$\mathfrak h=r^{-1}\rho(\partial_r^2-\partial_z^2)(r\psi)$ in
the \emph{non-orthogonal} skew coordinate $(r,\phi,z)$,
which may also be useful in some situations.

\subsection{Superintegrable cases}

Linear algebra dictates that if three functions $\mathfrak{s,h,k}$
are linearly dependent, the dimension of the linear space spanned
by the same three (i.e.\ the rank) is less than three.
Moreover, if the rank is one, then there exist two independent
combinations $(a,b,c)$ that result in equation (\ref{eq:cdep}),
which implies existence of two additional independent integrals of motion.
In other words, the potential is superintegrable and thus separable
in at least two different axisymmetric St\"ackel coordinates
if the rank of the set $\set{\mathfrak{s,h,k}}$ is one (or zero).

The rank of $\set{\mathfrak{s,h,k}}$ is one if all three
are constant (possibly zero) multiples of a common function.
The superintegrable potential with $\mathfrak k\ne0$ is separable
in the spheroidal or spherical coordinate with the origin at
$(x,y,z)=(0,0,\alpha)$ with an arbitrary $\alpha$ and
$\beta=\alpha^2-2\gamma\alpha+\tilde\beta
=(\alpha-\gamma)^2-\gamma^2+\tilde\beta$, where
$(\tilde\beta,2\gamma)=(\mathfrak s/\mathfrak k,\mathfrak h/\mathfrak k)$.
As before, the $\beta>0$, $\beta<0$, and $\beta=0$ cases correspond to
the separability in
the prolate and oblate spheroidal, and spherical polar coordinates,
respectively. As $\alpha\rightarrow\infty$, the coordinate tends to
the rotational parabolic coordinate with the origin at $(x,y,z)=(0,0,\gamma)$,
in which the potential is also separable. If $\mathfrak k=0$ on
the other hand, the potential is separable in the cylindrical polar
coordinate. In addition,
if $\mathfrak s/\mathfrak h=\tilde\alpha$ ($\mathfrak h\ne0$)
is also constant,
then $\mathfrak s-\tilde\alpha\mathfrak h-c\cdot0=0$ with
an arbitrary $c$, and so the potential is further separable in any spheroidal
and spherical coordinate with the origin at $(x,y,z)=(0,0,\tilde\alpha)$,
which basically corresponds to equation (\ref{eq:ham}) with a displaced origin.
The last remaining possibility of the superintegrable potential is
the $\mathfrak h=\mathfrak k=0$ case, which is discussed shortly.

As in the $\mathfrak s=\mathfrak k=0$ case (eq.~\ref{eq:ham}),
degenerate superintegrable cases that $\mathfrak s=\mathfrak h=0$ and
$\mathfrak h=\mathfrak k=0$ also completely specify
the axisymmetric potentials up to constant coefficients.
In particular,
\begin{align}
\mathfrak s&=\mathfrak h=0&\Leftrightarrow&&
\psi&=-k_0r^{-1}+(k_1+k_2\cos\theta)\,\rho^{-2}+k_3,
\label{eq:kep}\\\label{eq:pa-ca}
\mathfrak h&=\mathfrak k=0&\Leftrightarrow&&
\psi&=k_0(\rho^2+4z^2)+k_1\rho^{-2}+k_2z+k_3.
\end{align}
%
The potential in equation (\ref{eq:kep}) is separable in
the spherical polar and rotational parabolic
coordinates with the origin at $(0,0,0)$ as well as any prolate
spheroidal coordinate one of the foci of which is at $(0,0,0)$
(NB: $a\mathfrak s+b\mathfrak h+0\cdot\mathfrak k=0$ for any $a,b$,
and thus $\alpha=b/a$ is arbitrary and $\beta=\alpha^2\ge0$). If
$(\mathfrak s/\mathfrak k,\mathfrak h/\mathfrak k)=(\tilde\beta,2\gamma)$
is constant and $\tilde\beta=\gamma^2$, the corresponding
potential reduces to equation (\ref{eq:kep}) once the origin is
relocated to $(x,y,z)=(0,0,\gamma)$.
On the other hand, the potential in equation (\ref{eq:pa-ca})
is separable in the cylindrical polar coordinate
and any rotational parabolic coordinate with an arbitrary origin
on the symmetry axis ($2\gamma=\mathfrak h/\mathfrak k=0/0$
is indeterminate).
The superintegrability of both cases implies that
the meridional orbit projections are closed.
The maximal superintegrability is achieved 
for equation (\ref{eq:kep}) with $k_1=k_2=0$ (spherically symmetric)
or for equation (\ref{eq:pa-ca}) with $k_1=0$ (separable in
the Cartesian coordinate). Every orbit in either potential
is therefore all truly periodic.

The potential, $\psi=k_1\rho^{-2}+k_3$, is the intersection of
equations (\ref{eq:ham}), (\ref{eq:kep}) and (\ref{eq:pa-ca}),
and so $\mathfrak s=\mathfrak h=\mathfrak k=0$ (the null rank), indicating that
it is separable in \emph{any} axisymmetric St\"ackel coordinate.
Note that the meridional effective potential of an axially symmetric
potential $\psi$ is given by $\psi_\mathrm{eff}=\psi+(2\rho^2)^{-1}L_z^2$.
Here the centrifugal barrier is of this form, and thus
relevant mixed partial derivatives of $L_z^2/(2\rho^2)$ in
any axisymmetric St\"ackel coordinate also vanish. We infer that
an axially symmetric potential is separable in an axisymmetric coordinate
if the effective potential is separable in the meridional plane.

Equation (\ref{eq:kep}) with $k_1=k_2=0$ is the Kepler potential
($\psi=-k_0r^{-1}$).
In other words, the Kepler potential is separable
in the prolate spheroidal coordinate provided
that the point mass is located at either focus
defining the coordinate system (this reduces to the parabolic
coordinate as the other focus tends to the infinity). This
is already anticipated by the example in Sect.~\ref{sec:ex}
as the potential in equation (\ref{eq:Kd}) in the upper or
lower half is actually indistinguishable from that of a
point mass located on the symmetry axis in the opposite
side at the distance of $a$ from the mid-plane.
The classical consequences of this include that the gravitational potential
($\psi=-k_a\lVert\bmath r-\bmath a\rVert^{-1}
-k_b\lVert\bmath r-\bmath b\rVert^{-1}$)
due to two point masses (as well as the electric potential
of two point charges) is also in the St\"ackel form separable in the
prolate spheroidal coordinate with each mass lying at either focus
(\citealt{LL}, prob.~48-2; \citealt{Ar89}, sect.~47C).
In the limit when one of the point masses moves to the infinity, this results
in the $r^{-2}$-force plus an external uniform force field
(corresponding linear potentials), whose potential ($\psi=-k_0r^{-1}+k_2z$)
is separable in the rotational parabolic coordinate with the external
force acting along the symmetry axis and the coordinate origin at the
point mass location \citep[prob.~48-1]{LL}.

\section{Conclusion}
\label{sec:con}

Let us consider three vector-valued linear functions of a vector:
\begin{align}
\tens S(\bmath a)&=-\bmath r\cross(\bmath r\cross\bmath a)
=r^2\bigl[\bmath a-(\hat{\bmath r}\dotp\bmath a)\hat{\bmath r}\bigr];
\\\tens P(\bmath a)&=-\frac{\bmath r\cross(\hbk\cross\bmath a)
+\hbk\cross(\bmath r\cross\bmath a)}2
=z\bmath a-
\frac{(\bmath r\dotp\bmath a)\hbk+(\hbk\dotp\bmath a)\bmath r}2;
\\\tens Z(\bmath a)&=-\hbk\cross(\hbk\cross\bmath a)
=\bmath a-(\hbk\dotp\bmath a)\hbk.
\end{align}
The sufficient condition for an axisymmetric (about $\hbk$)
potential $\psi$ to admit the third integral of motion is that the vector fields
$\del\cross\bmath s=\he_\phi\mathfrak s$,
$\del\cross\bmath h=\he_\phi\mathfrak h$
and $\del\cross\bmath k=\he_\phi\mathfrak k$
(or $\mathfrak s$, $\mathfrak h$ and $\mathfrak k$)
are linearly dependent in the functional space.
Here $\bmath s=\tens S(\del\psi)$,
$\bmath h=\tens P(\del\psi)$ and $\bmath k=\tens Z(\del\psi)$.
That is, there exist three real \emph{constants} $a,b,c$ such that
$a\mathfrak s+b\mathfrak h+c\mathfrak k=0$ ($a^2+b^2+c^2\ne0$).
Equivalently we may say 
1) the null space of the set $\set{\mathfrak{s,h,k}}$ is non-trivial
(i.e.\ non-zero nullity), 2) the same set,
considered as elements in the functional space, is rank deficient
and 3) the dimension of
the linear space spanned by the same set is less than three, etc.
The third integral is given by $2I=aL^2+2b\omega+c\ell^2+2\xi$
or equivalently $2\tilde I=2(I-c\mathcal H)=aL^2+2b\omega-cv_z^2+2(\xi-c\psi)$.
Here $\xi$ is the solution of $\del\xi=a\bmath s+b\bmath h+c\bmath k$ whilst
the kinetic part of the integral consists of
the quadratic forms associated with the tensors $\tens S$, $\tens P$ and
$\tens Z$, namely $\bmath v\dotp\tens S(\bmath v)=L^2$,
$\bmath v\dotp\tens P(\bmath v)=\hbk\dotp(\bmath v\cross\bmath L)=\omega$,
and $\bmath v\dotp\tens Z(\bmath v)=v^2-v_z^2=\ell^2$.

The same condition is also necessary for the potential to be separable
in an axisymmetric St\"ackel coordinate and thus to admit an axisymmetric
quadratic integral in addition to the Hamiltonian and the axial component of
the angular momentum. The three functions $\mathfrak s$,
$\mathfrak h$ and $\mathfrak k$ reduce to simple mixed partial derivatives
of the potential in a particular coordinate, that is to say,
\begin{align}
\mathfrak s&=r^{-1}\partial_r\partial_\theta(r^2\psi)
&&\text{the spherical polar, $(r,\theta,\phi)$};
\\\mathfrak h&=r^{-1}\partial_\sigma\partial_\tau(r\psi)
&&\text{the rotational parabolic, $(\sigma,\tau,\phi)$};
\\\mathfrak k&=\partial_\rho\partial_z\psi
&&\text{the cylindrical polar, $(\rho,\phi,z)$},
\end{align}
where the parabolic coordinate in the meridional plane is
defined such that $(\sigma^2,\tau^2)=(r+z,r-z)$ and so $2r=\sigma^2+\tau^2$
($\mathfrak h$ transformed to the cylindrical or spherical polar coordinate
is found in eq.~\ref{eq:hfrak}).
In general, the potential is separable
in the cylindrical polar coordinate with the third integral
in the form of $2I_3=v_z^2+2g(z)$ if and only if $\mathfrak k=0$
(here $g'=\partial_z\psi$). On the other hand,
if $\mathfrak h/\mathfrak k=2\gamma$ is constant,
the potential is separable in the displaced rotational parabolic
coordinate with the origin at $(x,y,z)=(0,0,\gamma)$
with $I_3=\hbk\dotp(\bmath v\cross\bmath L_\gamma)+\xi$, where
$\bmath L_\gamma$ is the angular momentum about $z=\gamma$
and $\xi$ is the solution of $\del\xi=\bmath h-2\gamma\bmath k$.
The special case $\mathfrak h=0$ results in $\gamma=0$,
and thus the corresponding potential is separable
in the same rotational parabolic coordinate
used for expressing $\mathfrak h$ in the above.
Lastly, if $\mathfrak s=\alpha\mathfrak h+(\beta-\alpha^2)\mathfrak k$
for some constants $\alpha$ and $\beta$,
then the potential is separable in the prolate spheroidal ($\beta>0$),
the spherical polar ($\beta=0$) or the oblate spheroidal ($\beta<0$)
coordinate with the origin at $(x,y,z)=(0,0,\alpha)$.
The corresponding third integral may be expressed to be
$2I_3=L_\alpha^2+\beta v_z^2+2\xi$, where
$\del(\xi-\beta\psi)=\bmath s-\alpha\bmath h+(\alpha^2-\beta)\bmath k$.
This includes the case that $\mathfrak s/\mathfrak k=\beta$ is constant
(i.e.\ $\alpha=0$), for which the potential is, depending on $\beta$,
separable in the spheroidal or spherical (corresponding to the
$\mathfrak s=0$ case, for which $\alpha=\beta=0$) coordinate
with the origin at $(0,0,0)$.

If two amongst $\mathfrak{s,h,k}$ simultaneously vanish, then
the potential is superintegrable. In particular,
\begin{align}
\psi^{\mathrm H}&=k_0r^2+k_1\rho^{-2}+k_2z^{-2}+k_3
&\Leftrightarrow&&\mathfrak s=\mathfrak k=0;
\\\psi^{\mathrm K}&=-k_0r^{-1}+(k_1+k_2\cos\theta)\rho^{-2}+k_3
&\Leftrightarrow&&\mathfrak s=\mathfrak h=0;
\\\psi^{\mathrm A}&=k_0(\rho^2+4z^2)+k_1\rho^{-2}+k_2z+k_3
&\Leftrightarrow&&\mathfrak h=\mathfrak k=0.
\end{align}
In addition to two coordinate systems resulting in the vanishing
mixed partial derivatives, these are also separable in any (prolate or oblate)
spheroidal coordinate with the origin at $(0,0,0)$ for $\psi^\mathrm H$,
any prolate spheroidal coordinate with one focus at $(0,0,0)$
for $\psi^\mathrm K$ or any rotational parabolic coordinate with
an arbitrary origin on the symmetry axis for $\psi^\mathrm A$.
Note that additional superintegrability is also possible if
the tri-ratio $\mathfrak s:\mathfrak h:\mathfrak k$ is constant
(i.e.\ all three are constant -- including zero -- multiples of
a single function).

The necessity of these conditions for the separable potential
(and existence of the third integral) is the consequence
of the fact that the kinetic part of a quadratic integral of motion
(in the Euclidean space) is limited to be a particular type of functions.
In 
$\R^2$, this has been known
since 
\citet[see also \citealt{Wi37}, sect.~152]{Be57}. In particular,
the kinetic part of any integral of motion in $\R^2$ quadratic
to the momenta must be in the form of
$a_1v_x^2+2a_2v_xv_y+a_3v_y^2+a_4v_xL_z+a_5v_yL_z+a_6L_z^2$,
where $a_i$'s are constants and $L_z=xv_y-yv_x$.
In other words, the kinetic part is a
constant-coefficient degree-two homogeneous polynomial of
all the linear and angular momentum components.
This is also true in $\R^3$ (in fact, in $\R^n$).
If $p_i$ and $p_j$ are any two components of the linear or angular momenta,
then $p_ip_j$ forms the kinetic part of an integral, and so
the linear space spanned by all such $p_ip_j$
is a subspace of the kinetic part of all quadratic integrals.
Although there are 21 such factors
(i.e.\ 2-combination with repetition out of six, i.e.\ three linear
and three angular, elements), the dimension of this subspace is
in fact 20 thanks to $\bmath v\dotp\bmath L=0$.\footnote{The argument
generalizes for $\R^n$, i.e.\
there are $n$ linear and $n\choose2$ angular momentum components,
the sum of which corresponds to the dimension $m=\frac12n(n+1)$
of Euclidean isometry group $E(n)$.
This results in $\frac12m(m+1)=\frac18n(n+1)(n^2+n+2)$
quadratic `monomials', but they are not all independent
due to the 3-vector relation $\bmath v\wedge\mathsf J=0$,
which counts for $n\choose 3$ components,
and the further 4-vector one, $\mathsf J\wedge\mathsf J=0$,
which constitutes $n\choose 4$ linear constraints.
Here $\mathsf J=\bmath r\wedge\bmath v$ is the angular
momentum 2-vector. Hence, the vector space spanned by all quadratic
`monomials' is of the dimension $\frac1{12}n(n+1)^2(n+2)$.}
However, \citet{Ch39} had shown that
the kinetic part of a quadratic integral of motion in $\R^3$ contains
20 integration constant.\footnote{Note that $\sum_{i,j}\mathsf K_{ij}v_iv_j$
is the kinetic part of an integral of motion if and only if
$\mathsf K_{(ij;k)}=0$ (see Appendix~\ref{app}),
which reduces to $\left(\!\binom n3\!\right)$
homogeneous linear partial differential equations on
$\left(\!\binom n2\!\right)$ independent functions,
$\mathsf K_{ij}$ in any Cartesian coordinate of $\R^n$.
Here $\left(\!\binom nk\!\right)=\binom{n+k-1}k$ is
the $k$-combination with repetition out of $n$. In the Cartesian
coordinate, the covariant derivative reduces to the coordinate partial
derivative, which is symmetric for permuting indices.
Therefore, $\mathsf K_{(ij,k)lm}=0$, which results in
$\left(\!\binom n3\!\right)\cdot\left(\!\binom n2\!\right)$
homogeneous linear equations on the same number of independent functions,
$\mathsf K_{ij,klm}$, all of which identically vanish. Hence,
all $\mathsf K_{ij}$'s are quadratic polynomials of Cartesian
coordinate components whilst they are the simultaneous
solutions of $\mathsf K_{(ij,k)}=0$. Since
$\mathsf K_{(ij,k)l}=0$ are $q=\left(\!\binom n3\!\right)\cdot n$
linear equations on $p=\left(\!\binom n2\!\right)\cdot\left(\!\binom n2\!\right)$
independent functions, $\mathsf K_{ij,kl}$, integrating
$\mathsf K_{(ij,k)lm}=0$ introduces $p-q=\frac1{12}n^2(n+1)(n-1)$
constants of integration. Similarly, the next integration
of $\mathsf K_{(ij,k)l}=0$ introduces
$\left(\!\binom n2\!\right)\cdot n-\left(\!\binom n3\!\right)
=\frac13n(n+1)(n-1)$ whereas the final integration of
$\mathsf K_{(ij,k)}=0$ involves $\left(\!\binom n2\!\right)$
additional constants. Summing them up, the total number of
independent integration constants amounts to $\frac1{12}n(n+1)^2(n+2)$,
which is the same as the dimension of the linear space spanned
by quadratic monomials of the generators of $E(n)$.}
As the linear spaces of the same dimension,
the space of the kinetic part of quadratic integrals is thus
isomorphic to that of constant-coefficient degree-two homogeneous
polynomials of the linear or angular momentum
components,\footnote{That is, the space of the Killing
2-tensor is the grade-2 symmetric algebra over the space of
the Killing vectors, which are the generators of isometry.
The argument so far in the footnotes indicates that this is
true for any $\R^n$.}
and the kinetic part of any quadratic integral should be expressible
as one such polynomial.

Considering only axisymmetric kinetic parts 
leaves six independent bases, e.g.,
$\set{v^2,L^2,v_z^2,L_z^2,v_zL_z,\omega}$,
for quadratic polynomials. However, $v^2$ is the kinetic part of
$\mathcal H$ whereas $L_z$ (and so $L_z^2$) is the known integral
for any axially symmetric potential. Furthermore,
$D_t(v_zL_z)=\dot v_zL_z=-(\bmath L\dotp\hbk)\partial_z\psi
=\bmath v\dotp(\bmath r\cross\hbk\partial_z\psi)
=-\bmath v\dotp\he_\phi\partial_z(\rho\psi)$. That is,
if $v_zL_z+\xi$ is an integral, then $\del\xi=\he_\phi\partial_z(\rho\psi)$.
Restricting to axisymmetric integrals,
the only possibility is that the potential is translation invariant
as $\psi=\psi(\rho)$ and then $v_z$ becomes an integral,
which is again a trivial integral.
This leaves three degrees of freedom
(amongst these, one is simply the overall scale
whereas another can be subsumed into the choice of the origin)
for the kinetic part of the non-classical integral of motion.
Therefore, the assumed form $aL^2+2b\omega-cv_z^2$ is indeed the most general
form for the kinetic part of a nontrivial quadratic axisymmetric integral.

Traditionally, separable (St\"ackel) potentials are understood
to be given by the particular global functional forms in the special
set of coordinate systems whilst overlooking the fact that these forms
have originally been derived as general solutions to the set of partial
differential equations (relating to integrability conditions).
By focusing back on these underlying differential equations,
separable potentials are in principle characterizable in
any coordinate system and not just in the preferred coordinate system
in which the potential is expressible in the separable form. This is
important since, for a given potential, it is typically not known
a priori whether the special coordinate indeed exists and what
the particular coordinate is, even if it exists. We note
that existence of the third integral and regular orbits in an arbitrary
(not necessarily separable) potential may be understood by local
approximations to a separable potential \citep{ddZ88,Bi12,Sa12}.
Under this scenario, the ability of characterizing separable potentials
locally (via differential equations) is clearly crucial.

In recent time, after relative neglect \emph{post}
the influential works by Tim de~Zeeuw and collaborators,
astrophysical interests on the St\"ackel potential appear
to resurrect in light of efforts to construct
a three-integral distribution model of the Galaxy using the phase-space
action--angle coordinate \citep[e.g.,][]{Bi12,Sa12}.
The St\"ackel potentials are the most general class of the potentials
in which the actions can be calculated for all orbits via
`analytic' (up to matrix inversion and integral quadratures) means
\citep[see also Appendix~\ref{app:aav}]{Sa12}.
In order to find the transformation to the action integrals
$\set{\mathcal H,L_z,I_3}\to\set{J_1,J_2,J_\phi}$, it needs
to specify the frame set $\set{\he_1,\he_2,\he_\phi}$ --
and the orthogonal coordinate system $(q_1,q_2,\phi)$ --
in which the third integral
$I_3=\kappa_1v_1^2+\kappa_2v_2^2+\kappa_\phi v_\phi^2+\xi$ diagonalizes.
Assuming $I_3$ is independent from $\mathcal H$ and $L_z^2$
(i.e.\ $\kappa_1\ne\kappa_2$),
the squares of diagonalized velocity components are expressed
as a linear combination of three quadratic integrals,
\begin{equation}\label{eq:vso}\begin{split}
v_1^2&=\frac{(I_3-\xi)-2\kappa_2(\mathcal H-\psi)}{\kappa_1-\kappa_2}
-\frac{\kappa_3-\kappa_2}{\kappa_1-\kappa_2}\frac{L_z^2}{\rho^2};
\\v_2^2&=\frac{2\kappa_1(\mathcal H-\psi)-(I_3-\xi)}{\kappa_1-\kappa_2}
-\frac{\kappa_1-\kappa_3}{\kappa_1-\kappa_2}\frac{L_z^2}{\rho^2}
\end{split}\end{equation}
and $v_\phi^2=L_z^2/\rho^2$.
For an orthogonal frame, the coordinate scale factor can be
found using $h_i^{-2}=\lVert\del q_i\rVert^2$ (cf.\ $h_\phi=\rho$) and
the momentum component by $p_i=h_iv_i$ (cf.\ $p_\phi=\rho v_\phi=L_z$).
For a separable potential, $p_i^2=h_i^2v_i^2$ with $v_i^2$
in equation (\ref{eq:vso}) is then a function of $q_i$ alone.
Hence, the orbits in a separable potential are bounded by
the coordinate surfaces with accessible $q_i$
restricted to an interval where $p_i^2=h_i^2v_i^2\ge0$
for fixed values of the integrals (if the frequency
in each coordinate direction is independent from one another,
the bound orbit is dense in the bounded region).
The transform to the action on the other hand is given by
an integral quadrature,
\begin{equation}
2\pi J_i=\oint\rmd q_i\sqrt{h_i^2v_i^2}
\end{equation}
and $J_\phi=L_z$.
In practice however, if the St\"ackel coordinate
in which the potential separates is known, it is
straightforward to proceed with separation of variables
of the Hamilton--Jacobi equation, which results in
the expression for the momentum as a function of the separation
constants. Examples are provided in Appendix~\ref{app:aav}.

\small
\section*{acknowledgements}
The author appreciates Wyn Evans for his introduction to the subject,
and also acknowledges \citet{CZdZ},
the review of which was the earnest impetus for the project.
The author also thanks Stephen Justham and Gareth Kennedy for
reading the manuscript and Luca Naso for helping to translate \citet{LC04}.
This research has made use of {\sl Google Books} and
{\sl Google Books Library Project}, which provides
valuable services by making insights of great minds of the past
easily accessible to the wide community. The author is supported
by grants from the Chinese Academy of Sciences (CAS) and
the National Science Foundation of China (NSFC).

\normalsize
\begin{appendix}

\section{Separable potentials with quadratic integrals of motion}
\label{app}

If the motion of a tracer respects the same number of independent
integrals of motion (that are `in involution') as the degrees
of freedom (`Liouville-integrable'), the Liouville--Arnold theorem
\citep[see][sect.~49]{Ar89} indicates that its (bound and bounded) orbit
is characterized
to be a superposition of simple periodic motions in each degree of freedom.
The motion is referred to as conditionally periodic \citep{Ar89} or
quasi-periodic \citep{BT}, which results in a `regular' orbit.
It is thus of great theoretical importance to discover potentials
in three-dimensional space that admit at least three
integrals of motion, in which all orbits are regular.
Of particular interest amongst such are `separable' potentials
for which the Hamilton--Jacobi equation (HJE)
is soluble through additive separation of variables in a suitably
chosen coordinate \citep[see e.g.,][sect.~48]{LL}. Orbits in such separable
potentials are characterized by a filled subregion of the space bounded
by the level surfaces of the same coordinate provided that
periods of motion in each degree of freedom
are not commensurable. Notwithstanding C.~G.~J. Jacobi's
scepticism\footnote{\selectlanguage{german}{\glqq}Die
Hauptschwierigkeit bei der Integration gegebener Differentialgleichungen
scheint in der Einf\"uhrung der richtigen Variablen zu bestehen, zu deren
Auffindung es keine allgemeine Regel giebt. Man muss daher das
umgekehrte Verfahren einschlagen und nach erlangter Kenntnis einer
merkw\"urdigen Substitution die Probleme aufsuchen, bei welchen
dieselbe mit Gl\"uck zu brauchen ist.{\grqq} \citep[pp.198-199]{Jacobi}.
\selectlanguage{english}See \citet[p.266]{Ar89} for an English
translation.}, the separable potentials and the associated coordinates
with which the Hamilton--Jacobi method is applicable
have been completely characterized at least for natural dynamical systems.
Note that a natural dynamical system is characterized by
the Lagrangian of the form $\mathscr L=T-\psi$, where $T$ is a homogeneous
quadratic function of velocities and $\psi$ is the potential, which is
a scalar field on the space. Then
$p_i=\partial\mathscr L/\partial\dot q^i=\partial T/\partial\dot q^i$ and
$\sum_i\dot q^ip_i=\sum_i\dot q^i\partial T/\partial\dot q^i=2T$, and so
the Hamiltonian is given by $\mathcal H=\sum_i\dot q^ip_i-\mathscr L=T+\psi$.

It was Paul \citet{St91,St93} who had first shown that,
if the coordinates $(q^1,\dotsc,q^n)$ are orthogonal
(i.e.\ the metric is diagonalized)
such that the line element is given by $\rmd s^2=\sum_ih_i^2(\rmd q^i)^2$
and $2T=\sum_ih_i^2(\dot q^i)^2=\sum_ip_i^2/h_i^2$ (where
$p_i=h_i^2\dot q^i$ is the conjugate momentum to the coordinate $q^i$),
then the necessary and sufficient condition for the HJE to have a solution
whose dependences on different coordinates are separated (that is to say,
any mixed coordinate partial derivative vanishes)
is that (1) there exists a nonsingular (i.e.\ invertible) matrix of
functions $\set{\mathsf G_i^j(q^i)}$ such that
$\partial\mathsf G_i^j/\partial q^k=0$ for $k\ne i$ and
$\sum_i\mathsf G_i^j/h_i^2=\deltaup^j_1$ (where $\deltaup_i^j$ is
the Kronecker delta) -- which is equivalent to $h_i^2=G/C^1_i$, where
$C_i^j$ is the cofactor of the matrix $\set{\mathsf G_i^j}$
and $G=\det\set{\mathsf G_i^j}=\sum_i\mathsf G_i^jC_i^j$ is
its determinant (known as the St\"ackel determinant) --
and (2) the potential in the given coordinate system is in the form
of $2\psi=\sum_if_i(q^i)/h_i^2$, where $f_i(q^i)$ is an arbitrary function
of the sole coordinate component $q^i$. This is collectively
known as the St\"ackel condition \citep{Go80}.

The condition (1) does not involve the potential at all.
That is, separation of variables of the HJE in an orthogonal coordinate
is possible \emph{only if} the coordinate scale factors
$h_i$'s are given such that a particular matrix of functions
$\set{\mathsf G_i^j(q^i)}$ exists, irrespective of $\psi$.
Mildly abusing terminology, we refer to such orthogonal coordinates
as the `St\"ackel coordinates'. Note that
the St\"ackel coordinate is equivalent to the coordinate
in which the HJE of geodesic motions ($\psi=0$) is additively separable.
Somewhat confusingly, the potential as in the form of the condition (2)
is sometime referred to be in a separable form \emph{relative} to the specified
coordinate. By contrast, the separable potential itself is
defined absolutely such that there exists a St\"ackel coordinate
in which the potential is expressible in the separable form.
In this paper, to alleviate possible confusion,
the separable potential in the absolute sense
is referred to as the `St\"ackel potential' whereas the potential
is noted to be in the separable form if it is written down as
$2\psi=\sum_if_i(q^i)/h_i^2$ using some functions $f_i(q^i)$ and
the scale factors $h_i$'s of the orthogonal coordinate system
whose line element is $\rmd s^2=\sum_ih_i^2(\rmd q^i)^2$
(the metric is diagonalized, and $h_i^2$ and $h_i^{-2}$, respectively,
are the diagonal components of covariant and contravariant metric
coefficients).

The St\"ackel condition as given is operational;
once $\set{\mathsf G_i^j(q^i)}$ is known, it is trivial to solve
the HJE via separation of variables and to show that
$\alpha_j=\sum_i(p_i^2+f_i)\mathsf H^i_j$ is an integral of motion,
where $\set{\mathsf H^i_j(q^1,\dotsc,q^n)}$
is the inverse matrix of $\set{\mathsf G_i^j}$ (i.e.\
$\mathsf H^i_j=C_i^j/G$). However, it is fairly difficult
to verify directly whether the particular coordinate
is `St\"ackel' using this definition. For this purpose,
we refer to the insight of Tulio \citet{LC04}
who had realized that the solution of the HJE via separation of
variables is also the solution to the system of overdetermined
first-order quasi-linear partial differential equations, namely
$\partial p_i/\partial q_j=0$ for $i\ne j$ and
$\partial p_k/\partial q_k=-(\partial_k\mathcal H)/(\partial^k\mathcal H)$,
where $\mathcal H=\mathcal H(q^1,\dotsc,q^n;p_1,\dotsc,p_n)$ is
the Hamiltonian, and $\partial_k\mathcal H=\partial\mathcal H/\partial q^k$
and $\partial^k\mathcal H=\partial\mathcal H/\partial p_k$.
Thanks to the Frobenius integrability theorem,
the compatibility condition on the system -- i.e.\
$(\partial/\partial q_i)(\partial p_j/\partial q_j)
=(\partial/\partial q_j)(\partial p_j/\partial q_i)=0$ for $i\ne j$ --
is the necessary and sufficient condition for
the HJE to be solvable via separation of variables.
This is known as the Levi-Civita separability condition.

In a natural dynamical system, the Levi-Civita separability condition
reduces to $n\choose 2$ quartic even polynomial equations on $p_i$,
whose coefficients on each power of $p_i$ must identically vanish,
should the integrability condition be satisfied.
Whilst the zeroth coefficient becomes $g^{ij}\psi_{,i}\psi_{,j}=0$
(the subscript comma notation for the coordinate partial derivative)
for $i\ne j$, which is trivial in orthogonal coordinates,
the fourth and second coefficients in an orthogonal coordinate result
in the second-order partial differential equations on the metric coefficients
and the potential,
namely $\mathcal S_{jk}(h_i^{-2})=0$ and $\mathcal S_{jk}(\psi)=0$ with
$j\ne k$, where $\mathcal S_{jk}(f)=f_{,jk}
+(\ln h_k^2)_{,j}f_{,k}+(\ln h_j^2)_{,k}f_{,j}$.
Hence, the scale factors of the St\"ackel coordinate must be solutions
to $\mathcal S_{jk}(h_i^{-2})=0$ for $j\ne k$ whereas
the potential is in the separable form if it is the solution of
$\mathcal S_{jk}(\psi)=0$ for $j\ne k$ (NB: $\mathcal S_{jk}$ depends on
the chosen coordinate) and is `St\"ackel' if there exists a St\"ackel
coordinate in which $\mathcal S_{jk}(\psi)=0$ for all pairs $j$ and $k$
with $j\ne k$.

An alternative characterization of the St\"ackel condition follows the
observation that the integrals of motion of a natural dynamical system
resulting from separation of variables of the HJE are quadratic to $p_i$.
Consequently, existence of an integral of motion that is quadratic
to $p_i$ (other than the Hamiltonian) is in fact necessary for the HJE
to be soluble through separation of variables. In general,
the quadratic integral of motion is in the form of $I=\mathsf K^{ij}p_ip_j+\xi$
(the Einstein summation convention is used in this paragraph)
where $\mathsf K^{ij}$ is a symmetric tensor and $\xi$ is a scalar field.
Explicit calculations establish that $\dot I=
\mathsf K_{(ij;k)}\dot q^i\dot q^j\dot q^k+(\xi_{,i}
-2\mathsf K_i^j\psi_{,j})\dot q^i$
(where the semicolon and parentheses in the subscripts represent
the covariant derivative and the index symmetrization).
Hence, if one defines the vector-valued linear function of a vector
(i.e.\ 2-tensor) such that $\tens K(\bmath a)=\be_i\mathsf K^i_ja^j$
for $\bmath a=a^i\be_i$, where $\set{\be_i}$ is the coordinate basis,
then existence of $\tens K$ and $\xi$
such that $\mathsf K_{(ij;k)}=0$ and $\del\xi=2\tens K(\del\psi)$
is equivalent to $I=\tens K(\dot{\bmath r})\dotp\dot{\bmath r}+\xi$
being an integral, and also necessary for separation of variables of the HJE.
Generalizing the Killing (after Wilhelm Killing; 1847-1923) vector field $X_\mu$
for which $X_{(\mu;\nu)}=0$, the symmetric tensor $\mathsf K_{ij}$ such that
$\mathsf K_{(ij;k)}=0$ is referred to as the Killing 2-tensor.
Then, since $\del\wedge\del\xi=0$ (which is the same
as the Frobenius integrability condition), the condition of
$I$ being an integral is also equivalent to existence of the Killing
2-tensor\footnote{Any integral of motion that is an $n$th polynomial of $p_i$
implies existence of the Killing $n$-tensor such that
$\mathsf K_{(i_1\dotso i_n;i_0)}=0$. In particular, existence of
the Killing vector field $\bmath X$ such that $\bmath X\dotp\del\psi=0$
indicates that the natural Lagrangian is invariant along the integral
curve of $\bmath X$ and there exists an integral of motion which
is a liner combination of $p_i$ (the Noether theorem).}
$\tens K$ such that $\del\wedge\tens K(\del\psi)=0$, which also
characterizes the St\"ackel potential intrinsically
in a coordinate-free formulation.

The integrals due to separation of variables of the HJE in an orthogonal
coordinate have diagonalized quadratic terms in momenta, namely
$I=\alpha_j=\sum_i(p_i^2+f_i)\mathsf H^i_j$.
Consequently, only such Killing tensors that are globally diagonalizable
in an orthogonal coordinate are of relevance in the characterization
of the St\"ackel coordinate. Note the tensor $\tens K$ is globally
diagonalizable in an orthogonal coordinate if there exists a pair-wise
orthogonal foliation of space such that each leaf is normal to one of
the eigenvectors of $\tens K$ at every location.\footnote{This leads to
the Pfaffian system. Its integrability condition resulting from
the Frobenius theorem reduces to $\be\wedge(\del\wedge\be)=0$, which
is equivalent to $\be\dotp(\del\cross\be)=0$ in $\R^3$,
where $\be$ is an eigenvector of $\tens K$.}
In the orthogonal coordinate that diagonalizes the Killing tensor
such that $\mathsf K_i^j=\kappa_i\deltaup_i^j$ (no summation), the Killing
equation $\mathsf K_{(ij;k)}=0$ results in the system of first-order
linear partial differential equations on the set of eigenvalues,
namely $\partial\kappa_i/\partial q^j=
(\kappa_i-\kappa_j)(\partial\ln h_i^2/\partial q^j)$.
Then the integrability condition on existence of such $\kappa_i$'s -- i.e.\
$(\partial/\partial q^k)(\partial\kappa_i/\partial q^j)=
(\partial/\partial q^j)(\partial\kappa_i/\partial q^k)$ -- reduces
to $(\kappa_j-\kappa_k)h_i^2\mathcal S_{jk}(h_i^{-2})=0$ for $j\ne k$.
Hence, the orthogonal coordinate is `St\"ackel' if there exists
a Killing tensor that is diagonalizable in the given coordinate and
has all distinct eigenvalues \citep[cf.][]{Ei34}.
Conversely, in the St\"ackel coordinate, separation of variables of
the HJE leads to $n$ integrals of motion and thus there actually exist
$(n-1)$ independent Killing tensors (other than the metric) that
all diagonalize in the same St\"ackel coordinate.

This approach to the St\"ackel coordinates via the Killing tensor
provides its geometric interpretation. Instead of
algebro-analytic constraints, $\mathcal S_{jk}(h_i^{-2})=0$
for $j\ne k$, we find that the St\"ackel coordinate
surfaces must be the integral surfaces of particular differential systems
defined by the Killing tensors.
Further developments of the idea have led to intrinsic (coordinate-free)
characterization of separation of variables
\citep[see][]{Be80,Be97,KM80,KM81,KM82,KM86}.
An important result following this \citep{We24,Ei34,KM86,Be97} is that,
in the two or three-dimensional Euclidean (i.e.\ flat) space, the only
possible St\"ackel coordinates are the Jacobi elliptic/ellipsoidal coordinates
\citep[the 26th lecture]{Jacobi} and their degenerate forms\footnote{They
correspond to the elliptic, parabolic, polar, and Cartesian coordinates
in $\R^2$. In $\R^3$, they are the ellipsoidal, elliptic-paraboloidal, 
conical, prolate and oblate spheroidal, rotational-parabolic,
spherical-polar, elliptic-cylindrical, parabolic-cylindrical,
cylindrical-polar, and Cartesian coordinates. These 11 coordinates
are exactly the same in which the Helmholtz and also the
(time-independent) Schr\"odinger equation are solvable through
\emph{multiplicative} separation of variables. This is not a coincidence
since the \citet{Ro28} condition for separation of variables of
the Schr\"odinger equation consists of the St\"ackel condition
plus $(\prod_ih_i)/G$ being multiplicatively separable
where $G$ is the St\"ackel determinant \citep[see also][]{Ei34,MF53,KM80}.}
\citep{MF53} up to rotations and translations. This follows
the fact that the integral surfaces of the Killing tensor system are
confocal quadrics (or degenerate planes).

The equation on the potential $\del\wedge\tens K(\del\psi)=0$,
on the other hand reduces to $(\kappa_i-\kappa_j)\mathcal S_{ij}(\psi)=0$
for $i\ne j$ in the St\"ackel coordinate that diagonalizes
the given Killing tensor such that $\mathsf K_i^j=\kappa_i\deltaup_i^j$.
As noted earlier, the general solution of $\mathcal S_{ij}(\psi)=0$
is given by the potential in the separable form $2\psi=\sum_if_i(q_i)/h_i^2$
in the particular coordinate
\citep[e.g.,][see also \citealt{Wi37,Hi87}]{Da01}. Consequently,
existence of a quadratic integral of motion that is globally
diagonalizable with all distinct eigenvalues is also the sufficient condition
for the potential to be `St\"ackel' (and in the separable
form in the coordinate that diagonalizes the given quadratic integral)
and for the HJE to be soluble via separation of variables
(in the same coordinate that diagonalizes the integral). It also
follows that the given potential is `St\"ackel' if and only if
there exists a non-degenerately diagonalizable (i.e.\ with all distinct
eigenvalues) Killing 2-tensor such that $\del\wedge\tens K(\del\psi)=0$.
This last equation is also coordinate-independent and can thus specify
(the partial differential equation for) the St\"ackel potential
in an arbitrary coordinate once the coordinate components of
the Killing tensor are specified, the idea of which forms the basis
for the algorithmic test for the potential to be `St\"ackel' \citep{MW88,WRW03}.

In astrophysical contexts, the St\"ackel potential was first
introduced by \citet{Ed15}, who studied the potentials
consistent with the so-called Schwarzschild ellipsoidal hypothesis,
namely that the local velocity distribution of tracers in equilibrium is
in the form of ellipsoidal Gaussian (i.e.\ anisotropic Maxwellian
distributions). However, thanks to the \citet{Je15} theorem, the distribution
is an integral of motion if it is a solution to the collisionless Boltzmann
equation and therefore the ellipsoidal hypothesis in fact implies existence
of an integral $Q$ that is a quadratic function of velocities.
(The converse however is not true as the ellipsoidal hypothesis assumes
the specific one-integral distribution $f\propto\mathrm e^{-Q}$.)
In effect, \citet{Ed15} had assumed existence of a globally diagonalizable
Killing tensor and shown that the integrability condition on
the Killing equations implied that the orthogonal coordinate
that diagonalizes the Killing tensor must be characterized by confocal set
of quadric coordinate surfaces -- which had unbeknownst been proven
earlier by \citet{LC04}. Then \citet{Ed15} showed that, in the two- and
three-dimensional Euclidean space, the ellipsoidal hypothesis
(existence of a nondegenerate quadratic integral in actuality) further led
to the conclusion that the potential must be the separable form in
the same coordinate that diagonalizes the Killing tensor, which is
a suitably chosen ellipsoidal coordinate or its degenerate form
-- and thus the St\"ackel potential.

\citet{Ch39} had investigated the dynamics of stellar systems governed
by the one-integral distribution $f(Q)$ of a quadratic integral $Q$.
Although his discussion of velocity ellipsoids and associated potentials
was based on the nominally weaker assumption than that of \citet{Ed15},
the crux of the argument in actuality hinged on existence of a quadratic
integral $Q$. His approach was essentially the three-dimensional generalization of
that of \citet[see also \citealt{Wi37}, sect.~152]{Be57}. In terms of
the language here so far, he basically obtained 20 sets of differential
equations comprising ten defining the Killing 2-tensor $\mathsf K_{(ij;k)}=0$,
six for the Killing vector $X_{(i;j)}=0$, three corresponding to
the components of $\del\xi=2\tens K(\del\psi)$ and
one for $\bmath X\dotp\del\psi=0$.
In Euclidean space, if one adopts a Cartesian coordinate covering
the whole space, the 10 sets of the Killing equations for the 2-tensor
are integrated into quadratic functions of coordinates with 20 parameters
whilst those for the Killing vector into linear functions with 6
integration constants. By integrating the remaining differential
equations on the potential with these explicitly given Killing 2-tensor
and vector, he was then able to sort out the potentials
consistent with the assumption, the conclusion of which
was essentially identical to that of \citet{Ed15}
\citep[see however][sect.~3]{Ev11}.

However, the interest in the St\"ackel potentials
(then known as Eddington's potentials in astrophysical literature)
had remained limited until \citet{Ly62}
as they were believed not to be a good approximation
of `real' potential given the noticeable failure of the ellipsoidal
hypothesis, which had soon become clear in the period following
the initial work of \citet{Ed15}. On the other hand, \citet{Ly62} based
his work on assumption of existence of `local integral'. That is to say,
he assumed that there exists a fixed foliation in space such that
the family of potentials given by an arbitrary function on the foliation
all admits integrals of motion similarly through the variation of functions.
In natural dynamic systems, this then led to integrals
whose dependence on the momentum component normal to the foliation
leaf is only through the Hamiltonian. He then showed that this
implied that the dependence of the HJE on the coordinate components
tangential to the leaf is separated off. Consequently, his tabulation of
potentials admitting a `local integral' is equivalent to the list of
potentials with which the HJE is at least partially separable including
all the St\"ackel potentials (for which the HJE is \emph{completely} separable).
In addition, he also figured that in the given ellipsoidal coordinate,
three free functions $f_i$ in the separable potential
$2\psi=\sum_if_i(q^i)/h_i^2$ can be `glued' together to
a single real-valued $C^2$-function ($C^3$ if one forces the continuity
on the mass density) of one real variable.

This last fact was noted independently by \citet{Ku56},
who had investigated the mass profile generating the St\"ackel
potentials (see also \citealt{dZvdV} for discussion on the contribution
by Grigori G. Kuzmin to the subject). Note that the same fact implies
that the potential along the long-axis of the specified ellipsoidal
(or prolate spheroidal for axially symmetric potentials) coordinate
determines a unique St\"ackel potential separable in the given coordinate
-- that is, $\del\wedge\tens K(\del\psi)=0$ is uniquely integrated into
the particular solution $\psi$ given the one-dimensional boundary
condition specified along the preferred axis of the given Killing
tensor. \citet{Ku56} also showed that
the Laplacian of separable potentials
along the long axis of the prolate spheroidal (more generally, ellipsoidal)
coordinate was only related to the behaviour of the potential along the same axis
and subsequently the Poisson equation resulted in a linear second-order
\emph{ordinary} differential equation for the potential along the same axis
with the density along the same axis as the source term. From these, he
was able to derive the flattened three-dimensional mass profile
generating the St\"ackel potential separable in the prolate spheroidal
coordinate given arbitrary nonnegative -- which is sufficient for
the nonnegativity of the 3D profile \citep[`the Kuzmin--de~Zeeuw theorem';
see also][]{dZ85-kt} -- function for the density along
the short axis (which is actually the long axis of the coordinate surface).
Particularly notable amongst his models is
$\varrho(\rho,z)\propto(1+\rho^2/a^2+z^2/b^2)^{-2}$.
In fact, its triaxial generalizations
\citep[`the perfect ellipsoid'; see][]{dZ85-pe}
are the only ellipsoidally stratified density profile without singularity
that generate the St\"ackel potentials \citep{dZLB}.

\section{The Hamilton--Jacobi equation and action--angle coordinates}
\label{app:aav}

Suppose that the Hamiltonian is given in the form of
\begin{subequations}\begin{equation}
\mathcal H=\frac12\sum_{i=1}^n\frac{p_i^2}{h_i^2}+\psi(q^1,\dotsc,q^n).
\end{equation}
Then the Hamilton--Jacobi equation (HJE) is the partial differential
equation for the Hamilton principal function $S(q^1,\dotsc,q^n;t)$,
\begin{equation}\textstyle
\frac12\sum_{i=1}^nh_i^{-2}(\partial_iS)^2
+\psi+\partial_tS=0.
\end{equation}
Since $\partial_t\mathcal H=0$, we may instead consider
the reduced HJE
\begin{equation}\textstyle
\sum_{i=1}^nh_i^{-2}(W_{,i})^2
+2(\psi-E)=0
\end{equation}\end{subequations}
by setting $S=W(q^1,\dotsc,q^n)-Et$, where $E$ is a constant.

Assuming that
the chosen orthogonal coordinate satisfies the St\"ackel condition,
that is, there exists an invertible ($n\times n$)-matrix of
functions $\set{\mathsf G^j_i(q^i)}$ such that
\begin{subequations}\begin{equation}\label{eq:stkm}
\sum_{i=1}^n\frac{\mathsf G_i^j(q^i)}{h_i^2}=
\begin{cases}1&(j=1)\\0&(j=2,\dotsc,n)\end{cases},
\end{equation}
%
and that the potential is of the separable form
\begin{equation}\label{eq:stkp}
\psi=\sum_{i=1}^n\frac{f_i(q^i)}{2h_i^2}
\end{equation}
in the same coordinate,
the HJE further reduces to
\begin{equation}
\sum_{i=1}^n\frac1{h_i^2}\left[
\left(\frac{\partial W}{\partial q^i}\right)^2+f_i(q^i)
-\sum_{j=1}^n\alpha_j\mathsf G_i^j(q^i)\right]=0,
\end{equation}
where $\set{\alpha_1=2E,\dotsc,\alpha_n}$ are a set of $n$ constants.
Hence, if
\begin{equation}\label{eq:dwdq}\textstyle
w_i(q^i)=\int\!p_i(q^i)\,\rmd q^i
\,;\quad
(p_i)^2=\sum_{j=1}^n\alpha_j\mathsf G_i^j(q^i)-f_i(q^i),
\end{equation}
then
$S=\sum_{i=1}^nw_i(q^i;\alpha_i,\dots,\alpha_n)-\alpha_1t/2$
is the complete solution of the HJE.
The canonical transform given by the generating function
$S=\sum_{i=1}^nw_i-Et$ leaves the transformed Hamiltonian
identically zero, and thus every $\alpha_i$ is conserved.
The expression for $\alpha_i$ in the old coordinate
is found from $p_i=\partial_iS=w_i'$,
\begin{equation}\label{eq:hjint}\textstyle
\alpha_j=\sum_{i=1}^n[p_i^2+f_i(q^i)]\,\mathsf H^i_j(q^1,\dotsc,q^n),
\end{equation}\end{subequations}
where $\set{\mathsf H^i_j(q^1,\dotsc,q^n)}$ is
the inverse matrix of $\set{\mathsf G_i^j}$. Hence, $\alpha_i$ is
an \emph{isolating} integral of motion.
They are also in involutions and therefore
the dynamic system is Liouville integrable.
The natural phase-space coordinate for such a
system is the action--angle coordinate, in which
the bound orbit specified by the level surface of the full set
of $n$ isolating integrals is equivalent to
$n$-torus embedded in the $2n$-dimensional phase space.

The action variable is formally defined to be
\begin{subequations}\begin{equation}
J_i(\alpha_1,\dotsc,\alpha_n)=
\frac1{2\pi}\oint_{\gamma_i}\textstyle{\sum_{j=1}^n}p_j\,\rmd q^j,
\end{equation}
where $\set{\gamma_1,\dotsc,\gamma_n}$ is the set of cycles
that forms the basis of the orbital torus (i.e.\ the level set
of the $n$ isolating integrals).
If the integrals are the separation constants of the HJE as in
equation (\ref{eq:hjint}), we can choose the cycles such that
along $\gamma_i$, only $q^i$ varies and all other $q^j$ ($j\ne i$)
are held fixed. The action variable is then given by
an integral quadrature,\footnote{Here we have assumed that the motion
in the $q^i$ direction is basically an oscillation between $q^i_{\min}$
and $q^i_{\max}$. Depending on the precise nature of the actual motion
in the $q^i$ coordinate, the integral limits and the multiplicative factor
accounting for symmetry must be chosen appropriately instead.}
\begin{equation}\label{eq:act}
J_i=\frac1{2\pi}\oint\rmd q^ip_i
=\frac1\pi\!\int_{q^i_{\min}}^{q^i_{\max}}\rmd q^ip_i(q^i),
\end{equation}
where the integral is over the interval $q^i\in[q^i_{\min},q^i_{\max}]$
in which $p_i^2\ge0$ (eq.~\ref{eq:dwdq}).
The complete set of $n$ equations for $J_i(\alpha_1,\dotsc,\alpha_n)$
then defines the transformation
$(\alpha_1,\dotsc,\alpha_n)\to(J_1,\dotsc,J_n)$.

The conjugate set of the angle variables is found using
the generating function $W=\sum_{i=1}^nw_i$ with $\alpha_i$
given by the inverse functions $\alpha_i=\alpha_i(J_1,\dotsc,J_n)$.
That is to say,
\begin{equation}\label{eq:ang}
\vartheta^i=\frac{\partial W}{\partial J_i}
=\sum_{j,k=1}^n\mathsf A_j^i\frac{\partial w_k}{\partial\alpha_j},
\quad
\mathsf A_j^i=\frac{\partial\alpha_j}{\partial J_i},
\end{equation}
where $\mathsf A^i_j$ is the Jacobian matrix element of
$\alpha_i(J_1,\dotsc,J_n)$. Strictly
$\mathsf A^i_j$' is a function of $J_i$, but with equation (\ref{eq:act})
they can also be considered as functions of $\alpha_i$. Then
since $\alpha_i(J_1,\dotsc,J_n)$ is
the inverse of $J_i(\alpha_1,\dotsc,\alpha_n)$,
the Jacobian matrix of $J_i(\alpha_1,\dotsc,\alpha_n)$
is the inverse matrix of $(\mathsf A^i_j)$. That is, $\mathsf A^i_j$
as a function of $\alpha_i$ may be found using
$\sum_{k=1}^n\mathsf J^i_k\mathsf A^k_j=
\sum_{k=1}^n\mathsf A^i_k\mathsf J^k_j=\deltaup^i_j$, where
\begin{equation}
\mathsf J_i^j(\alpha_1,\dotsc,\alpha_n)
=\frac{\partial J_i}{\partial\alpha_j}
=\frac1{2\pi}\oint\rmd q^i\frac{\mathsf G^j_i(q^i)}{2p_i(q^i)}.
\end{equation}
In addition, we infer from equation (\ref{eq:dwdq}) that
\begin{equation}
\frac{\partial w_k}{\partial\alpha_j}
=\int\!\rmd q^k\frac{\mathsf G^j_k(q^k)}{2p_k(q^k)},
\end{equation}
which is a function of $q^k$ and the parameter set
$\set{\alpha_1,\dotsc,\alpha_n}$. Equation (\ref{eq:ang}) therefore
provides with us the transformation
$(q^1,\dotsc,q^n;\alpha_1,\dotsc,\alpha_n)\to(\vartheta^1,\dotsc,\vartheta^n)$.
Finally, the canonical transformation to the action--angle coordinate
$(q^1,\dotsc,q^n;p_1,\dotsc,p_n)\to
(\vartheta^1,\dotsc,\vartheta^n;J_1,\dotsc,J_n)$.
is then given by combining equation (\ref{eq:ang}) with 
equations (\ref{eq:hjint}) and (\ref{eq:act}). Note that
the resulting transformation only involves analytic operations
up to integral quadratures (i.e.\ simple antiderivatives)
and matrix inversions.

Since the generating function of the canonical transform
to the action--angle coordinate does not explicitly involve the time,
the Hamiltonian in the action--angle coordinate is simply
$\mathcal H=E=\alpha_1(J_1,\dots,J_n)/2$, and the Hamilton
equations of motion are
\begin{equation}
\dot J_i=-\frac{\partial\mathcal H}{\partial\vartheta^i}=0,\quad
\dot\vartheta^i=\frac{\partial\mathcal H}{\partial J_i}
=\frac12\frac{\partial\alpha_1}{\partial J_i}.
\end{equation}\end{subequations}
Hence, the angle variable $\vartheta^i$ evolves linearly with time
in the constant angular frequency $\Omega^i(J_1,\dots,J_n)=\mathsf A^i_1/2$.
The frequency as a function of the orbital torus defined by
the integrals $(\alpha_1,\dotsc,\alpha_n)$ is again found by
the matrix element of the inverse matrix of $(\mathsf J_i^j)$ -- for $n=2,3$,
this is easily found using the cofactors.

\subsection{Axisymmetric St\"ackel potentials}

For any axisymmetric potential, the axial component of the angular
momentum $L_z=p_\phi$ is also the action integral:
\begin{equation}
J_\phi=\frac1{2\pi}\oint\rmd\phi\,p_\phi
=\frac1{2\pi}\int_0^{2\pi}\!\rmd\phi\,L_z=L_z.
\end{equation}

If the potential is separable in the cylindrical polar coordinate,
$\psi=f(\rho)+g(z)$ (eq.~\ref{eq:cyl}),
then $E_z=\frac12v_z^2+g(z)$ is the independent integral
whilst the momentum is separated as in
\begin{equation}
p_\rho^2=2\bigl[E-E_z-f(\rho)\bigr]-\rho^{-2}L_z^2;\quad
p_z^2=2\bigl[E_z-g(z)\bigr].
\end{equation}

For the potential $\psi=f(r)+r^{-2}g(\theta)$
separable in the spherical polar coordinate (eq.~\ref{eq:sph}),
$I_3=\frac12L^2+g(\theta)$ is the third integral.
Then the separated momentum components are (NB.\ $p_\theta=rv_\theta$)
\begin{equation}
p_r^2=2\bigl[E-r^{-2}I_3-f(r)\bigr];\quad
p_\theta^2=2\bigl[I_3-g(\theta)\bigr]-L_z^2\csc^2\!\theta
\end{equation}

In the rotational parabolic coordinate $(\sigma,\tau,\phi)$
defined such that $\sigma^2=r+z$ and $\tau^2=r-z$, the general
separable potential is in the form of equation (\ref{eq:ppara})
with the third integral given by $I_3=\hbk\dotp(\bmath v\cross\bmath L)+\xi$
with $\xi$ in equation (\ref{eq:xpsep}). With the coordinate scale
factors $h_\sigma^2=h_\tau^2=\sigma^2+\tau^2=2r$, we then find
\begin{equation}\begin{split}
p_\sigma^2&=2rv_\sigma^2=
2[\sigma^2E-I_3-\tilde f(\sigma)]-\sigma^{-2}L_z^2;\\
p_\tau^2&=2rv_\tau^2=
2[\tau^2E+I_3-\tilde g(\tau)]-\tau^{-2}L_z^2.
\end{split}\end{equation}

As for the case that the potential is separable in a spheroidal coordinate,
we need to define a specific coordinate system. Here we consider
the coordinate variables given by two solutions $\chi$ of
\begin{equation}
\chi^{-1}\rho^2+(\chi+\beta)^{-1}z^2=1,
\end{equation}
which we set $\lambda\ge\mu$.
The inverse transform $(\lambda,\mu)\to(\rho^2,z^2)$
is given by equation (\ref{eq:sphct}) with $\alpha=0$
whilst the coordinate scale factors are
\begin{equation}
h_\lambda^2=\frac{\lambda-\mu}{4\lambda(\lambda+\beta)};\quad
h_\mu^2=\frac{\mu-\lambda}{4\mu(\mu+\beta)}.
\end{equation}
This is the prolate coordinate if $\beta=a^2>0$
for which $\lambda\ge0\ge\mu\ge{-\beta}$,
and the oblate coordinate if $-\beta=a^2>0$
for which $\lambda\ge{-\beta}\ge\mu\ge0$.
The general separable potential is then given by (eq.~\ref{eq:stk}) 
\begin{equation}
\psi=\frac{f(\lambda)-f(\mu)}{\lambda-\mu}
\end{equation}
with the corresponding third integral in the form of
\begin{equation}
I_3=\frac{L^2+\beta v_z^2}2
+\frac{(\mu+\beta)f(\lambda)-(\lambda+\beta)f(\mu)}{\lambda-\mu}.
\end{equation}
Finally, the momentum component as a function of its conjugate coordinate
is found to be
\begin{equation}\begin{split}
4\lambda(\lambda+\beta)p_\lambda^2&=(\lambda-\mu)v_\lambda^2=
2\bigl[(\lambda+\beta)E-\!I_3\!-\!f(\lambda)\bigr]-\beta\lambda^{-1}\!L_z^2;
\\
4\mu(\mu+\beta)p_\mu^2&=(\mu-\lambda)v_\mu^2=
2\bigl[(\mu+\beta)E-\!I_3\!-\!f(\mu)\bigr]-\beta\mu^{-1}\!L_z^2.
\end{split}\end{equation}

The action integral $J_i$ corresponding to the coordinate $q^i$
can now be found using equation (\ref{eq:act}), which then results in
the transformation $J_1(E,I_3,L_z)$ and $J_2(E,I_3,L_z)$.
Next since $p_\phi=L_z=J_\phi$, we have $w_\phi(\phi)=p_\phi\phi=J_\phi\phi$,
and so the generating function is written down as
$W=w_1(q^1;E,I_3,L_z)+w_2(q^2;E,I_3,L_z)+J_\phi\phi$ with
$w_i$ given by equation (\ref{eq:dwdq}).
The angle variables are then found using
\begin{equation}\begin{split}
\vartheta^1&=\frac1{D_1}\left(
\frac{\partial J_2}{\partial I_3}\frac{\partial\tilde w}{\partial E}
-\frac{\partial J_2}{\partial E}\frac{\partial\tilde w}{\partial I_3}
\right);
\\\vartheta^2&=\frac1{D_1}\left(
\frac{\partial J_1}{\partial E}\frac{\partial\tilde w}{\partial I_3}
-\frac{\partial J_1}{\partial I_3}\frac{\partial\tilde w}{\partial E}
\right);
\\\vartheta^\phi&=\phi+\frac1{D_1}\left(
D_2\frac{\partial\tilde w}{\partial E}
+D_3\frac{\partial\tilde w}{\partial I_3}
\right)+\frac{\partial\tilde w}{\partial L_z},
\end{split}\end{equation}
where $\tilde w(q^1,q^2;E,I_3,L_z)=w_1+w_2$ and
\begin{equation}\begin{split}
D_1&=\frac{\partial J_1}{\partial E}
\frac{\partial J_2}{\partial I_3}
-\frac{\partial J_2}{\partial E}
\frac{\partial J_1}{\partial I_3},
\\D_2&=\frac{\partial J_1}{\partial I_3}
\frac{\partial J_2}{\partial L_z}
-\frac{\partial J_2}{\partial I_3}
\frac{\partial J_1}{\partial L_z},
\\D_3&=\frac{\partial J_1}{\partial L_z}
\frac{\partial J_2}{\partial E}
-\frac{\partial J_2}{\partial L_z}
\frac{\partial J_1}{\partial E}.
\end{split}\end{equation}
Given $\mathcal H=E(J_1,J_2,J_\phi)$,
the angular frequency $\Omega^i(E,I_3,L_z)$ of the evolution
for $\vartheta^i$ is given by
\begin{equation}\begin{split}
\Omega^1=\dot\vartheta^1&=\frac{\partial\mathcal H}{\partial J_1}
=\frac1{D_1}\frac{\partial J_2}{\partial I_3}
=\left(\frac{\partial J_1}{\partial E}-\frac{\partial J_2}{\partial E}
\frac{\partial J_1/\partial I_3}{\partial J_2/\partial I_3}\right)^{-1},
\\\Omega^2=\dot\vartheta^2&=\frac{\partial\mathcal H}{\partial J_2}
=-\frac1{D_1}\frac{\partial J_1}{\partial I_3}
=\left(\frac{\partial J_2}{\partial E}-\frac{\partial J_1}{\partial E}
\frac{\partial J_2/\partial I_3}{\partial J_1/\partial I_3}\right)^{-1},
\\\Omega^\phi=\dot\vartheta^\phi&=\frac{\partial\mathcal H}{\partial J_\phi}
=\frac{D_2}{D_1}.
\end{split}\end{equation}

\end{appendix}
\label{lastpage}
\end{document}